\documentclass[twocolumn,aps,superscriptaddress,multicol,amsmath,amssymb]{revtex4-1}
\usepackage[utf8]{inputenc}
\usepackage{amsmath}
\usepackage{siunitx}
\usepackage{graphicx}
\usepackage{sidecap}
\usepackage[dvipsnames]{xcolor}
\graphicspath{{.}{./figs/}}
\usepackage{nameref,zref-xr}
\zxrsetup{toltxlabel}
\usepackage[colorlinks,linkcolor=blue,citecolor=blue,anchorcolor=blue]{hyperref}
\usepackage{cleveref}
\zexternaldocument*[SI-]{SI}
\usepackage {times}

\hypersetup{colorlinks=true,citecolor=blue,linkcolor=red,urlcolor=blue}
\usepackage{hyperref,cleveref}
\graphicspath{{Figs/}} 
\begin{document}
\title{Strain-induced stripe phase in charge ordered single layer NbSe$_2$}
\author{Fabrizio Cossu}
\email{fabrizio.cossu@apctp.org}
\affiliation{Asia Pacific Center for Theoretical Physics, Pohang, 37673, Korea}
\author{Kriszti\'an Palot\'as}
\affiliation{Institute for Solid State Physics and Optics, Wigner Research Center for Physics, H-1525 Budapest, Hungary}
\affiliation{MTA-SZTE Reaction Kinetics and Surface Chemistry Research Group, University of Szeged, H-6720 Szeged, Hungary}
\author{Sagar Sarkar}
\affiliation{Asia Pacific Center for Theoretical Physics, Pohang, 37673, Korea}
\author{Igor Di Marco}
\affiliation{Asia Pacific Center for Theoretical Physics, Pohang, 37673, Korea}
\affiliation{Department of Physics and Astronomy, Uppsala University, Box 516, SE-75120, Uppsala, Sweden}
\affiliation{Department of Physics, POSTECH, Pohang, 37673, Korea}
\author{Alireza Akbari}
\email{alireza@apctp.org}
\affiliation{Asia Pacific Center for Theoretical Physics, Pohang, 37673, Korea}
\affiliation{Department of Physics, POSTECH, Pohang, 37673, Korea}
\affiliation{Max Planck POSTECH Center for Complex Phase Materials, POSTECH, Pohang 790-784, Korea}
\affiliation{Max Planck Institute for the Chemical Physics of Solids, D-01187 Dresden, Germany}
\date{\today}
\begin{abstract}
%
  Charge density waves are ubiquitous phenomena in metallic transition metal dichalcogenides. In NbSe$_2$, a triangular $3\times3$ structural modulation is coupled to a charge
 modulation. Recent experiments reported evidence for a triangular-stripe transition at the surface, due to strain or accidental doping and associated to a $4\times4$ modulation.
 We employ \textit{ab-initio} calculations to investigate the strain-induced structural instabilities in a pristine single layer and analyse the energy hierarchy of the structural
 and charge modulations. Our results support the observation of phase separation between triangular and stripe phases in 1H-NbSe$_2$, relating the stripe
 phase to compressive isotropic strain, favouring the $4\times4$ modulation. The observed wavelength of the charge modulation is also reproduced with a good accuracy.
\end{abstract}
\maketitle

\section{Introduction}

  Transition metal dichalcogenides are among the most exciting hosts for ordered phases, such as superconductivity and charge/spin
 modulations~\cite{saito-NatRevMat2016,wilson-PRL.32.882,wilson-AdvPhys2001,moncton-PRL.34.734,revolinsky-JPandCS1965} because of
 their cooperative interactions~\cite{kiss-nphys2007}. In fact, their coexistence is counter-intuitive, being collective electronic
 modes which build an excitation gap, hence requiring a large number of electrons available at the Fermi level. Since early
 works~\cite{rice-PRL.35.120,liu_r-PRL.80.5762,straub-PRL.82.4504}, the transition to a charge/spin ordered state driven by the electron
 gas at the Fermi level has been questioned, partly because of the difficulties in explaining the intricate coexistence of superconductivity
 with charge density waves
 (CDWs)~\cite{wilson-PRL.32.882,moncton-PRL.34.734,revolinsky-JPandCS1965,rahn-PRB.85.224532,castroneto-PRL.86.4382,cho-ncomm2018}. Among
 the proposed explanations, the electronically driven mechanism was ruled
 out~\cite{inosov-NJP2008,rossnagel-PRB.64.235119,rossnagel-PRB.76.073102,johannes-PRB.73.205102,johannes-PRB.77.165135,zhu-PNAS2015,silvaguillen-2DMat2016}
 in favor of a momentum-assisted
 mechanism~\cite{calandra-PRB.80.241108,valla-PRL.92.086401,rahn-PRB.85.224532,zheng-PRB.97.081101,arguello-PRB.89.235115,flicker-ncomm2015,zhu-PNAS2015},
 which is likely influenced by external conditions, such as applied fields~\cite{galvis-ncommPhys2018,cho-ncomm2018} or epitaxially induced
 strain~\cite{fu_zg-EPL2017,wei_mj-PRB.96.165404}, and
 chemical~\cite{disalvo-PRB.12.2220,chatterjee-ncomm2015,cossu-PRB.98.195419} or gate~\cite{wei_mj-PRB.96.165404,shao-PRB.94.125126}
 doping.

  Further intricacies in the occurrence of these collective electronic excitations arise at the surface and single layers due to a reduced symmetry. While the bulk, 2H type, is
 characterised by a $P6_{3}/mmc$ symmetry, with centrosymmetry and inversion symmetry operations, the layer at the surface has no inversion symmetry and lacks a van der Waals bonded
 layer; its structural type remains 2H, although high reactivity to (accidental) dopants and diffusion into interstitial regions affect experimental investigations, modifying its
 structure and symmetry operations; the single layer is characterised by a $D_{3h}$ symmetry, with an inversion symmetry operation but lack of centrosymmetry and no van der Waals
 bonded layers on either sides. In passing from the bulk to thin layers and down to a single layer, enhanced effects from fluctuations and lack of van der Waals bonded layers may
 determine dramatic changes in the structural and electronic degrees of
 freedom~\cite{kotov-RMP.84.1067,guinea-ANDP2014,mak-nmat2012,frindt-PRL.28.299,xi-nnano2015,ugeda-nphys2016,butler-ACSNano2013,geim-Nature2013,novoselov-Science2016,ge-PRB.86.104101}.
 Due to the fact that CDW consists of periodic lattice distortions coupled to charge modulations, structural properties may modify such collective excitation dramatically, and in turn
 its interplay with superconductivity~\cite{menard-nphys2015,ugeda-nphys2016}. In the case of 2H-NbSe$_2$, the CDW is recognised to arise from a phonon
 instability~\cite{weber_f-PRL.107.107403}, inducing an electronic reconstruction in a sizable energy range~\cite{arguello-PRB.89.235115} in the bulk and at the surface, although single
 layers show a CDW gap centred at the Fermi level~\cite{ugeda-nphys2016}. Electronic and magnetic texture may be layer-resolved~\cite{riley-nphys2014,bawden-ncomm2016}, contributing to
 a complex scenario to explain the thickness-dependent properties of these materials. Reduced dimensionality allows for different ways of tuning the crystal and the electronic structure.
 The competition between the trigonal prismatic (2H) and the octahedral (1T) phases becomes stronger; as a consequence, the 1T crystal phase, unstable in the bulk~\cite{kadijk-JTLCM1971},
 may be synthesised~\cite{nakata-NPGAM2016} or induced by external sources~\cite{bischoff-ACSchemat2017}; the resulting charge order is not the usual star of David phase, unlike
 TaSe$_2$~\cite{liu_y-PRB.94.045131,boerner-APL2018} and TaS$_2$~\cite{law-PNAS2017,fazekas-PhysBC}. In the 2H-type, $4\times4$~\cite{soumyanarayanan-PNAS2013,gao-PNAS2018} and/or
 $2\times2$~\cite{gao-PNAS2018} modulations, linked to a stripe phase at the surface, are reported. Distinct CDW structures, supporting the widely known CDW incommensurate character in
 2H-NbSe$_2$~\cite{gye_gc-PRL.122.016403} are observed, further speculating on their topological connection. A change from the $3\times3$ to the $4\times4$ modulation was previously
 predicted by \textit{ab-initio} studies on 2H/1H-type NbSe$_{2}$~\cite{calandra-PRB.80.241108,lian-NanoL2018.5}, but an early scanning tunnelling microscopy/spectroscopy (STM/STS)
 study~\cite{ugeda-nphys2016} confirms that the $3\times3$ modulation dominates in single layers deposited on bilayer graphene; on the other hand, a recent theoretical
 study~\cite{flicker-PRB.92.201103} showed that within a $3\times3$ modulation, uniaxial strain leads to a triangular-stripe CDW transition in 2H-NbSe$_{2}$. However, phase transitions
 remain often latent because they are suppressed by growth/synthesis conditions, choice of the substrate and/or quantum fluctuations. In this respect, discerning intrinsic from extrinsic
 characteristics is crucial to understand these phases and finding possible routes to their manipulation.

  The present work focusses on 1H-NbSe$_2$ single layers under isotropic biaxial strain. In order to shed light on the competition and coexistence of
 various periodicities and structures, we first perform phonon calculations, suggesting the wave-vectors of the structural instabilities. Thereafter,
 the suggested periodicities are investigated with a full set of total energy calculations, yielding both the energy hierarchy of the possible CDW
 structures and their charge distributions. A $3{\mathbf q}$-$1{\mathbf q}$ (triangular-stripe) transition occurs in the $4\times4$ modulation, whereas
 it is incipient in the $3\times3$ modulation. Although our model represents only an approximation of detailed experimental conditions, e.g. due to the
 fact that we do not include a substrate, we believe it identifies the most important intrinsic characteristics of single layers under isotropic biaxial
 strain. Moreover, our results are likely applicable also to thin  films, due to the fact that CDW properties depend on the structural properties more
 strongly than on the electronic properties, as well as due  to the above mentioned high reactivity to accidental doping which may affect the structural
 features at the surface.
 
\begin{figure*}[t]
\centering
 \includegraphics[trim = 0 0 0 0,width=0.9\textwidth,clip]{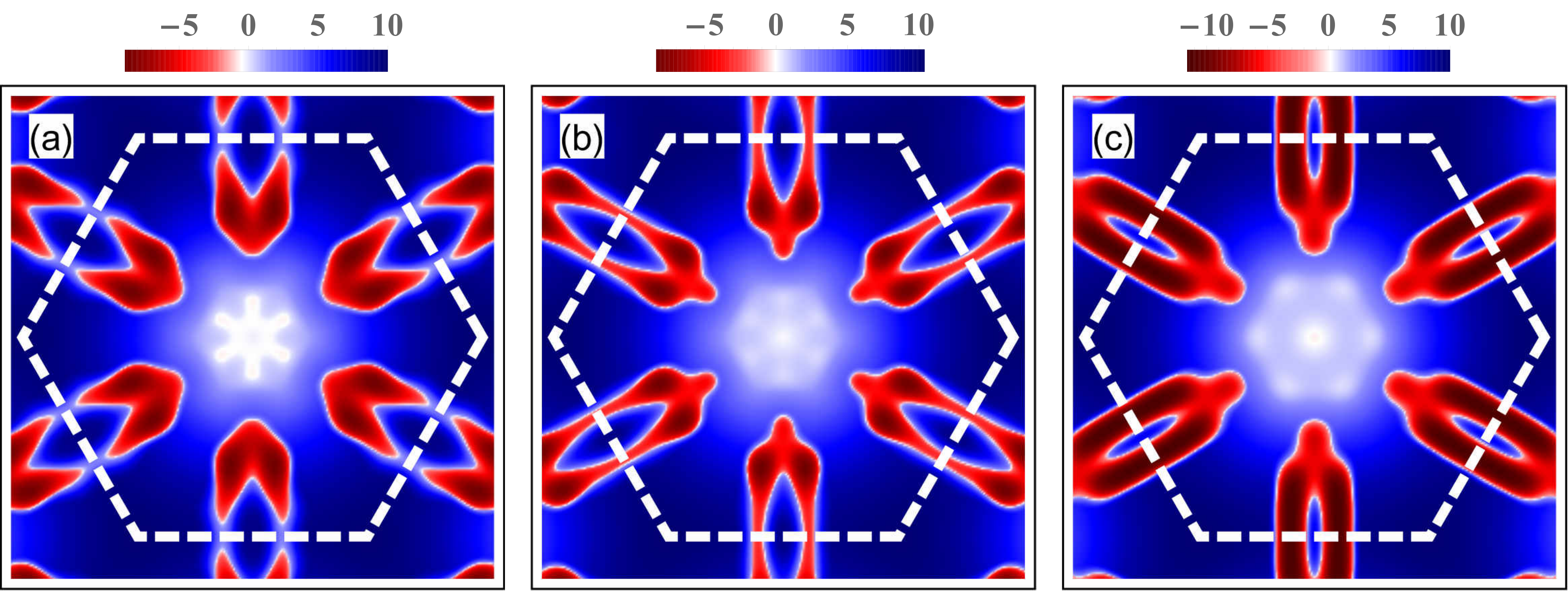}
 \caption{(Colour online) Two-dimensional phonon dispersion for the NbSe$_2$ single
 layer unit cell as two-dimensional density plots, under compressive strain (a),
 at the equilibrium lattice constant (b) and under tensile strain (c). The plots,
 as function of $k_x$ and $k_y$, in units of $2\pi/a$, show only the branch with
 the lowest values of $\omega({\mathbf q})^2$; red (blue) represent regions of
 negative (positive) values of $\omega({\mathbf q})^2$, see the colour bar, where
 the scale is given in tens of \SI{}{\per\cm}.}
\label{phonon-fig}
\end{figure*}

\section{Methods}


\begin{table*}[bth]
\centering
 \caption{Total energy of various CDW structures, grouped per periodicity, at
 three different values of the in-plane lattice constant. The energy of each
 structure is given as the difference, expressed in \si{\meV}/f.u., with the
 energy of the symmetric structure, and it is negative for favoured structures.
 The names of the structures are explained as follows: MM and MW in the
 $2\times3$ cells are two stripe phases with a different Nb-Nb pattern; HX, CC
 and HC in the $3\times3$ cells are the known hexagonal, triangular chalcogen
 centred and triangular hollow centred structures, respectively; CCccws, HXHC
 and HC in the $\sqrt{13}\times\sqrt{13}$ cells are phases with triangular
 chalcogen centred plus counterclockwise vortex, hexagonal with triangular
 chalcogen centred merged and just triangular chalcogen centred, respectively;
 finally, the $1{\mathbf q}'$, $3{\mathbf q}$ and $1{\mathbf q}''$ in the
 $4\times4$ cells are a single ordering vector CDW structure, a triple ordering
 vector CDW structure and a single ordering vector structure, respectively, as
 also illustrated in Fig.~\ref{4x4strctchg-fig}.}
 \label{tab-energy}
 \begin{tabular}{r||c|c|c|c|c|c|c|c|c|c|c}
                 & \multicolumn{2}{c|}{$2\times3$} & \multicolumn{3}{c|}{$3\times3$} & \multicolumn{3}{c|}{$\sqrt{13}\times\sqrt{13}$} & \multicolumn{3}{c}{$4\times4$} \\
    \hline
                        &  MM  &  MW  &  HX  &  CC  &  HC  &CCccws& HXHC &  HC  & $1{\mathbf q}'$ & $3{\mathbf q}$ & $1{\mathbf q}''$ \\
  $|\mathbf{a}|$ = 3.49 &$-1.9$&$-0.9$&$-3.0$&$-3.5$&$-4.4$&$-0.3$&$-0.4$&$-0.4$&     $-2.9$      &     $ -- $      &     $-3.0$      \\
  $|\mathbf{a}|$ = 3.45 &$-1.9$&$-0.0$&$-2.8$&$-3.4$&$-4.0$&$-1.4$&$-0.6$&$-1.4$&     $-4.4$      &     $-2.5$      &     $-4.3$      \\
  $|\mathbf{a}|$ = 3.41 &$-2.3$&$-0.0$&$-2.8$&$-3.6$&$-3.6$&$-3.4$&$-2.4$&$-3.6$&     $-6.2$      &     $ -- $      &     $-6.2$      \\
 \end{tabular}
\end{table*}


  Our results are obtained by {\itshape{ab initio}} calculations within the formalism of the density-functional theory (DFT). The projected augmented wave (PAW) method with
 Perdew-Burke-Ernzerhof (PBE) pseudopotentials \cite{blochl-PRB.50.17953,kresse-PRB.59.1758}, as implemented in the {\sc Quantum ESPRESSO} suite~\cite{giannozzi-JPCM2009,giannozzi-JPCM2017}
 and the Vienna Ab-initio Simulation Package (VASP), is used. The relaxation of the unit cell and the computation of its relative phononic spectra are run in the former code,
 while structural relaxation, total energy calculations and charge distributions are performed with the latter. The exchange-correlation functional is treated within the
 generalised gradient approximation using the PBE parametrisation~\cite{perdew-PRL.77.3865,perdew-PRLerratum.78.1396}. The pseudopotentials used in the simulations provide
 explicit treatment of the following valence electrons: $4s^{2}4p^{6}5s^{1}5d^{4}$ for Nb and $4s^{2}4p^{4}$ for Se. The cutoff energy of the plane waves for the unit cell
 (\textsc{Quantum ESPRESSO}) is $\sim$~\SI{1142}{\electronvolt} and that for the supercell calculations is \SI{500}{\electronvolt}; prior to the phonon calculations, an energy
 cutoff is also applied for the charge density and potential, of $\sim$~\SI{4354}{\electronvolt}. The energy tolerance on the electronic loops for the calculations in the unit
 cell is \SI{1.4e-9}{\electronvolt}, whereas it is \SI{e-6}{\electronvolt} for the calculations in the supercells (\SI{e-7}{\electronvolt} for accurate determination of the
 electronic properties, such as density of electronic states and charge distributions). A conjugate gradient algorithm is employed for structural relaxation in all cases, with
 a tolerance in energy of \SI{e-4}{\electronvolt} per unit cell and a residual force of $\sim$~\SI{e-3}{\electronvolt\per\angstrom} on each atom. The calculations of phonons
 are performed sampling the first Brillouin Zone with a $45\times45\times1$ k-grid (for the wavefunctions) and a $15\times15\times1$ q-grid (for the force constants). The formation
 of CDWs is analysed sampling the first Brillouin Zone by means of the $30\times30\times1$, $23\times23\times1$, $20\times20\times1$, $17\times17\times1$ and $15\times15\times1$
 grids of ${\mathbf k}$-points for the $2\time2\times1$, $\sqrt{7}\times\sqrt{7}\times1$, $3\times3\times1$, $\sqrt{13}\times\sqrt{13}\times1$ and $4\times4\times1$ supercells,
 respectively, keeping approximately an equal distance between ${\mathbf k}$-points.
 
  Finally, we also computed some quantities offering a more direct comparison to experimental data. Band structures were calculated using method of Popescu and
 Zunger~\cite{popescu-PRB.85.085201}. STM simulations were also performed, by means of the BSKAN code~\cite{hofer-PSS2003,palotas-JPCM2005}. The revised Chen
 method~\cite{mandi-PRB.91.165406} with an electronically flat and spatially spherical tip orbital has been used, which is equivalent to the Tersoff-Hamann model
 of electron tunnelling~\cite{tersoff-PRB.31.805}. The reported STM images are in constant current mode (with the maxima of the current contours at 5.8 \AA for
 all structures for a better comparability). Additional technical details on band structures and STM spectra are provided in the Appendix.

\section{Results and discussion}
\begin{figure*}[tbh]
\centering
 \includegraphics[trim = 0 0 0 0,width=0.95\textwidth,clip]{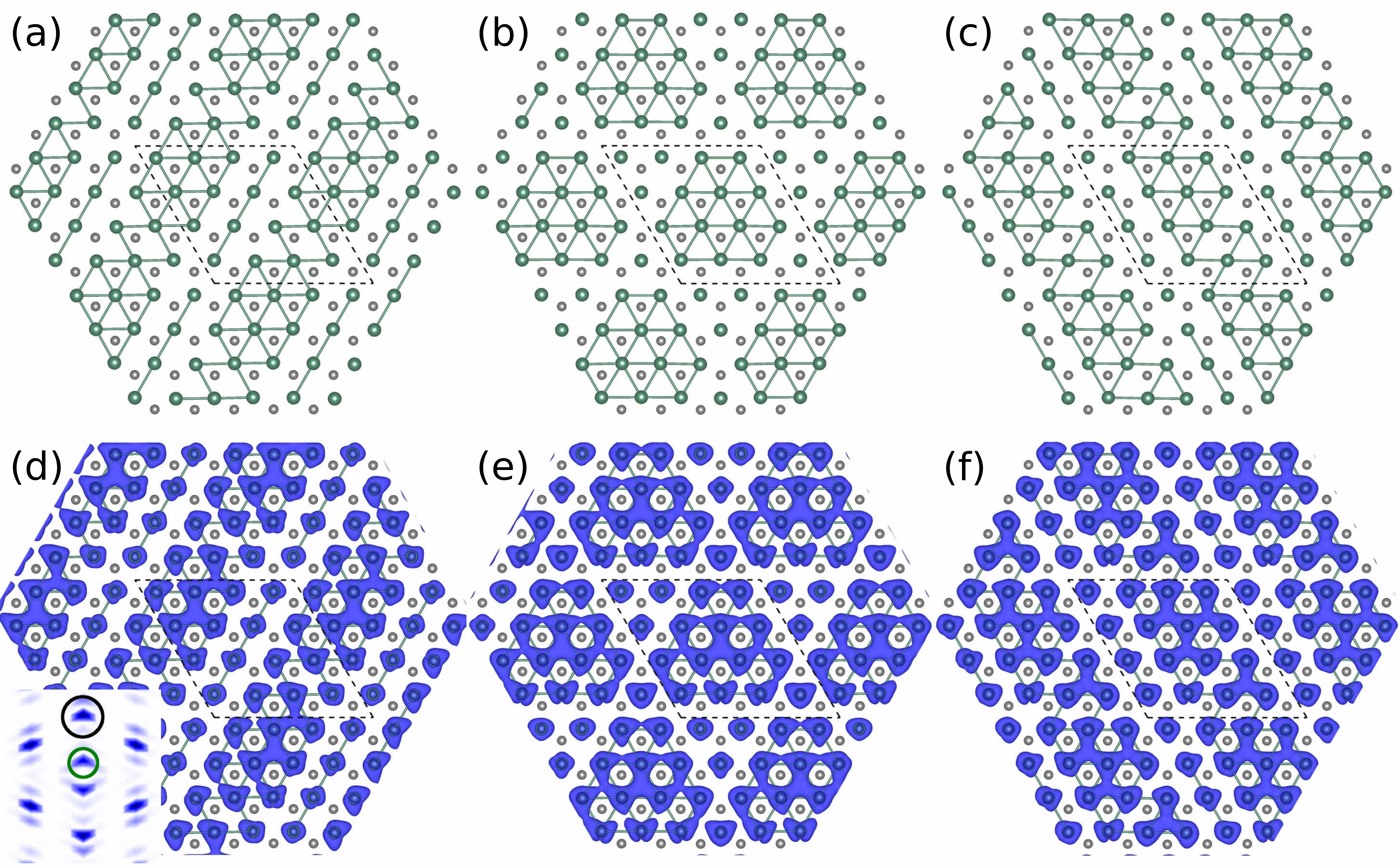}
 \caption{Structures (a-c) and corresponding charge (d-f) modulations
 resulting from a $4\times4$ periodicity. The wavelength of the stripe
 phases $1{\mathbf q}'$ -- (a) and (d) -- and $1{\mathbf q}''$ -- (c)
 and (f) -- is \SI{11.8}{\angstrom}; the central column -- (b) and (e)
 -- refers to the $3{\mathbf q}$ phase. Dark green (grey) spheres represent
 Nb (Se) atoms; Nb-Nb bonds shorter than the equilibrium distance (the
 respective lattice constant) are represented by solid lines, in order
 to help visualising the CDW structure pattern. Dashed lines mark the
 supercells borders. The inset in (d) shows the Fourier transform of
 the charge density of the $1{\mathbf q}'$ phase as computed in a
 $4\times4\sqrt{3}$ rectangular cell; the black (green) circle denotes
 one of the Bragg (CDW) peaks.}
 \label{4x4strctchg-fig}
\end{figure*}

  Structural relaxation of the unit cell was performed for three values of the lattice constant, \SI{3.41}{\angstrom}, \SI{3.45}{\angstrom} and \SI{3.49}{\angstrom},
 which aim at modelling the system under compressive strain ($\sim 1\%$), no strain and tensile strain ($\sim 1\%$), respectively. The precise value of the computed
 in-plane stress is \SI{7.2e-3}{\electronvolt\per\cubic\angstrom}, \SI{1.1e-3}{\electronvolt\per\cubic\angstrom} and \SI{-4.5e-3}{\electronvolt\per\cubic\angstrom},
 respectively. The small residual compressive strain present in the no-strain case is due to the choice of a standard literature value for the equilibrium lattice
 constant of the bulk, in line with the work of Fang {\em et al.}~\cite{fang-ScienceAdv2018}. Since the discrepancy is very small, this detail does not affect the
 conclusions of the present work. A recent \textit{ab-initio} study~\cite{silvaguillen-2DMat2016} reports a lattice constant of \SI{3.458}{\angstrom}, in agreement
 with the residual strain found for $\left|{\mathbf a}\right| = $ \SI{3.45}{\angstrom} in our calculations. The relaxed structures were used to compute phonon spectra,
 whose imaginary values identify regions of structural instability, thus speculating on the periodicity of the supercells describing the stable crystal. These supercells
 are then investigated by total energy calculations to confirm (or dismiss) the relevance of a periodicity and to evaluate the energy hierarchy of a set of symmetries
 within each periodicity. The lattice reconstructions considered on the basis of the analysis of phonon spectra are of the type $\sqrt{n}\times\sqrt{n}$~\cite{kamil-JPCM2018}
 and $n\times m$ (suggested by discussion of results from STM~\cite{gye_gc-PRL.122.016403}).

  The phonon dispersions are represented as two-dimensional plots, Fig.~\ref{phonon-fig}, with positive (negative) values of $(\omega({\mathbf q}))^2$)
 shown in blue (red). Compressive strain, equilibrium, and tensile strain are shown from left to right, (a) to (c). The deepest troughs of the single
 layer phonon dispersion appear off the $\Gamma$M line, as a consequence of the non-centrosymmetric character of the single layer (1H). The instabilities
 have to be interpreted as perturbations which drive the system away from its most symmetric configuration, providing only an indication, rather than a
 determination, of the resulting periodicities. A $\sqrt{13}\times\sqrt{13}$ modulation is expected from the lowest values of the phonon dispersion, see
 especially Fig.~\ref{phonon-fig} (b), where the troughs appear in couples at symmetric positions with respect to the $\Gamma$M high symmetry line. Each
 couple of instability represent two periodic lattice distortions with opposite chirality. This result supports early~\cite{komori-JPSJ1997} and
 recent~\cite{wang_hui-JPCM2009} STM observations. In particular, a 2H-1T structural phase separation is reported in refs.\ \onlinecite{wang_hui-JPCM2009}
 and \onlinecite{bischoff-ACSchemat2017,nakata-NPGAM2016}. Further, we note similarities between the phonon dispersion in the 1H-type and that in the
 1T-type~\cite{kamil-JPCM2018}. Nevertheless, these instabilities may couple to give rise to a $3\times3$ modulation, which is the most observed. In a
 similar fashion, instability regions around M may couple to give rise to a $2\times2$ (or even a $2\times3$) modulation. Instabilities are expected to
 dominate along $\Gamma$M in a 2H-type symmetry (because the crystal is overall centrosymmetric).

  Compressive strain reduces the instabilities along the MK line and the minima of $(\omega({\mathbf q}))^2$ dominate around 2/4 $\Gamma$M (off the high symmetry line), see
 Fig.~\ref{phonon-fig} (a), pointing towards a $4\times4$ modulation. Conversely, a zone of instability centered around, but not including, M becomes dominant under tensile
 strain, pointing towards a $2\times2$ modulation; further local minima of $(\omega({\mathbf q}))^2$ appear at 2/4 $\Gamma$M, Fig.~\ref{phonon-fig} (c), justifying recent
 reports~\cite{gao-PNAS2018} for which both $2\times2$ and $4\times4$ periodicities arise under tensile strain.

  After having found the most likely periodicities for the occurrence of periodic lattice distortions by the calculation of phonon spectra,
 we investigate a set of structures for each periodicity by total energy calculation; in particular, the distorted structures we start with
 have fewer point symmetry operations with respect to the full symmetric structure, and model a periodic lattice distortion with three-fold
 rotational symmetry. In order to keep a close connection to existing materials, we also explore periodicities compatible with experimental
 findings \cite{gao-PNAS2018,gye_gc-PRL.122.016403}. We study a representative number of structures for each periodicity (in the $3\times3$
 periodicity we study three known CDW structures~\cite{cossu-PRB.98.195419,gye_gc-PRL.122.016403,malliakas-JACS2013,lian-NanoL2018.5} anew,
 complementing previous works by showing the energy hierarchy upon strain); here we refer to them as hexagonal (HX), chalcogen centered
 triangular (CC) and hollow centered triangular (HC). Energy differences, shown in Table~\ref{tab-energy}, are used to test whether and how
 each periodicity can give rise to a CDW structure.

\begin{figure}[b]
\centering
 \includegraphics[trim = 0 0 0 0,width=0.48\textwidth,clip]{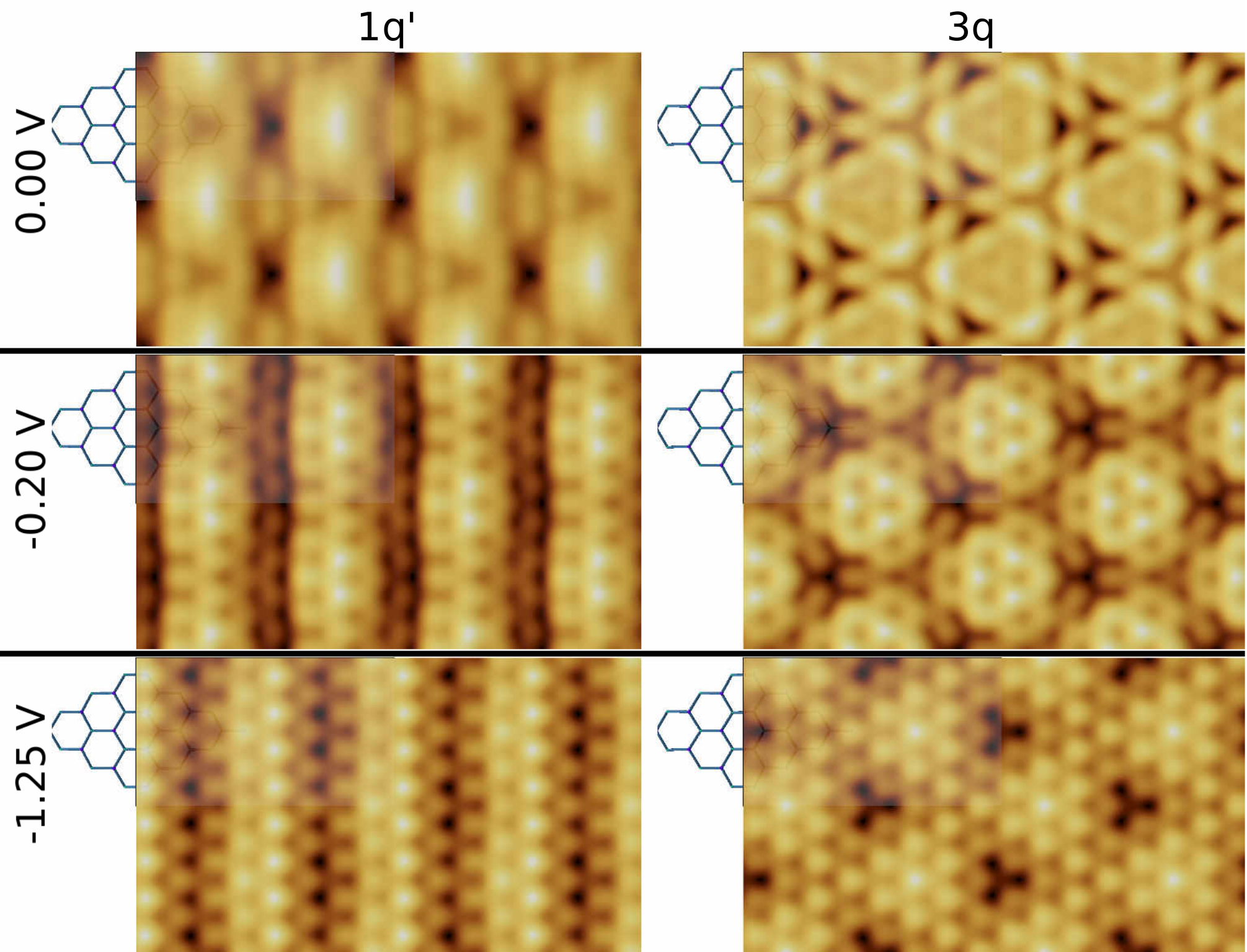}
 \caption{STM maps for $1{\mathbf q}'$ and $3{\mathbf q}$ $4\times4$ CDWs,
 left and right column, respectively; raws 1, 2 and 3 stand for \SI{0.00}{V},
 \SI{-0.20}{V} and \SI{-1.25}{V}, respectively.}
\label{staps4x4-fig}
\end{figure}

  All $2\times2$ structures tested in our calculations relaxed to the symmetric structure, for all values of the strain; therefore, the $2\times2$ periodicity in ref.\
 \onlinecite{gao-PNAS2018} does not arise from an intrinsic character of a single layer. The $2\times3$ periodicity accounts for the occurrence of instabilities at 2/3
 $\Gamma$M and at M; compare with ref.\ \onlinecite{gye_gc-PRL.122.016403}, figure 3, where HC and CC merge. The energy hierarchy between CDW structures with $3\times3$
 periodicity is in agreement with existing \textit{ab-initio} studies~\cite{lian-NanoL2018.5,cossu-PRB.98.195419}. The $\sqrt{13}\times\sqrt{13}$ and the $4\times4$
 periodicities allow the relaxation of three structures each, energetically close to the known $3\times3$ CDWs. In particular, the $4\times4$ CDW structures are favoured
 over the $3\times3$ CDW ground state under compressive strain. In general, a plethora of structural modulations are available in single layer NbSe$_2$, whose competition
 depends and may be tuned by epitaxial strain, charge transfer or proximity effects.

\begin{figure}[t]
\centering
 \includegraphics[trim = 0 218 0 20,width=0.48\textwidth,clip]{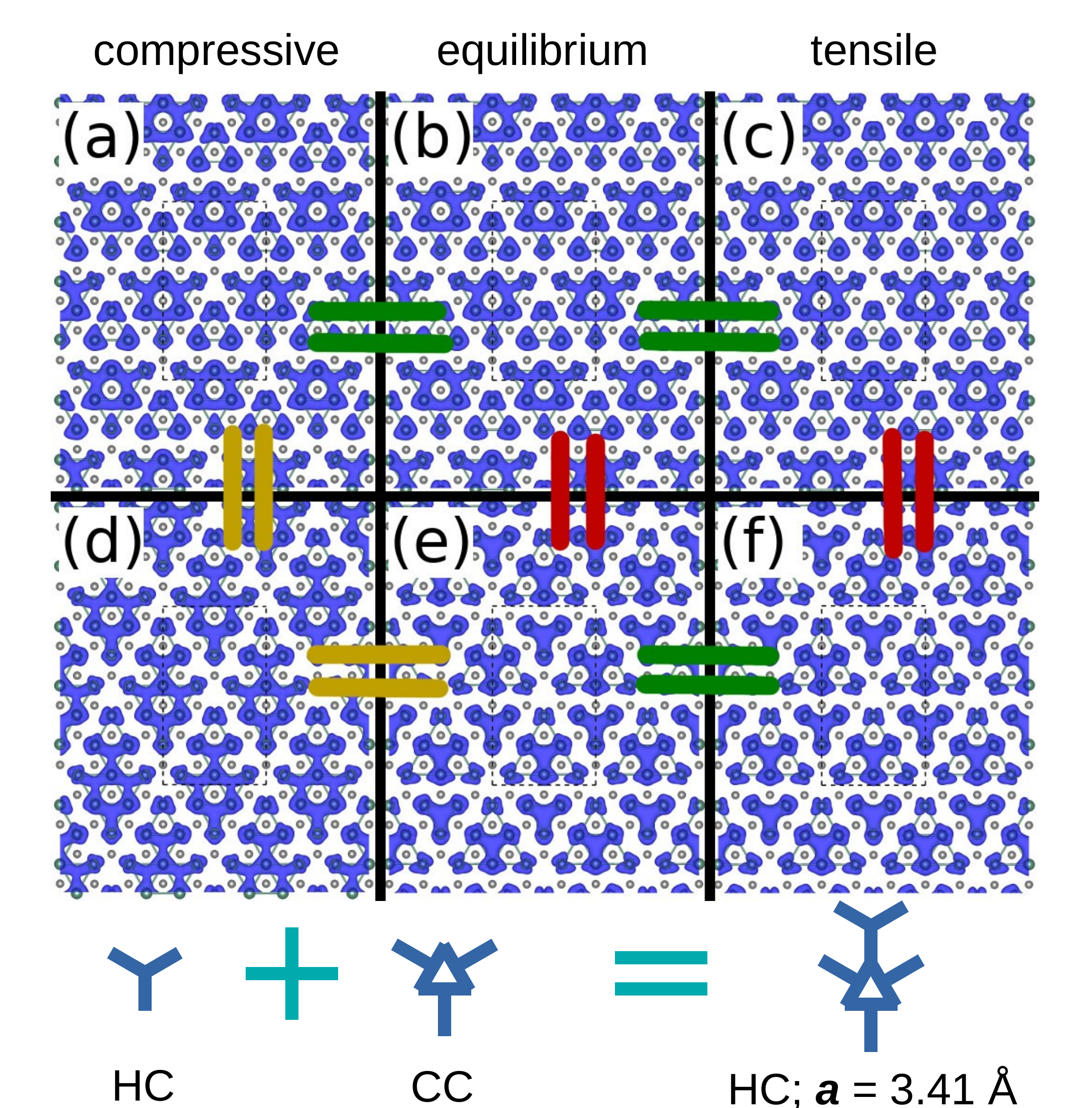}
 \caption{Charge distribution of the CC (a)-(c) and the HC (d)-(f) CDW
 structures under compressive strain (a,d), no strain (b,e) and tensile
 strain (c,f). At the bottom, a sketch shows how the charge patches in
 the HC and CC merge in the HC CDW under compressive strain, as appears
 in (d). Each pair of green, yellow and red bars connect two structures
 with equivalent, similar and different charge patches, respectively. At
 the bottom, a sketch shows what happens in figure (d), as two different
 structures merge under compressive strain.}
\label{3x3striperect-fig}
\end{figure}

\begin{figure}[t]
\centering
 \includegraphics[trim = 0 0 0 0,width=0.45\textwidth,clip]{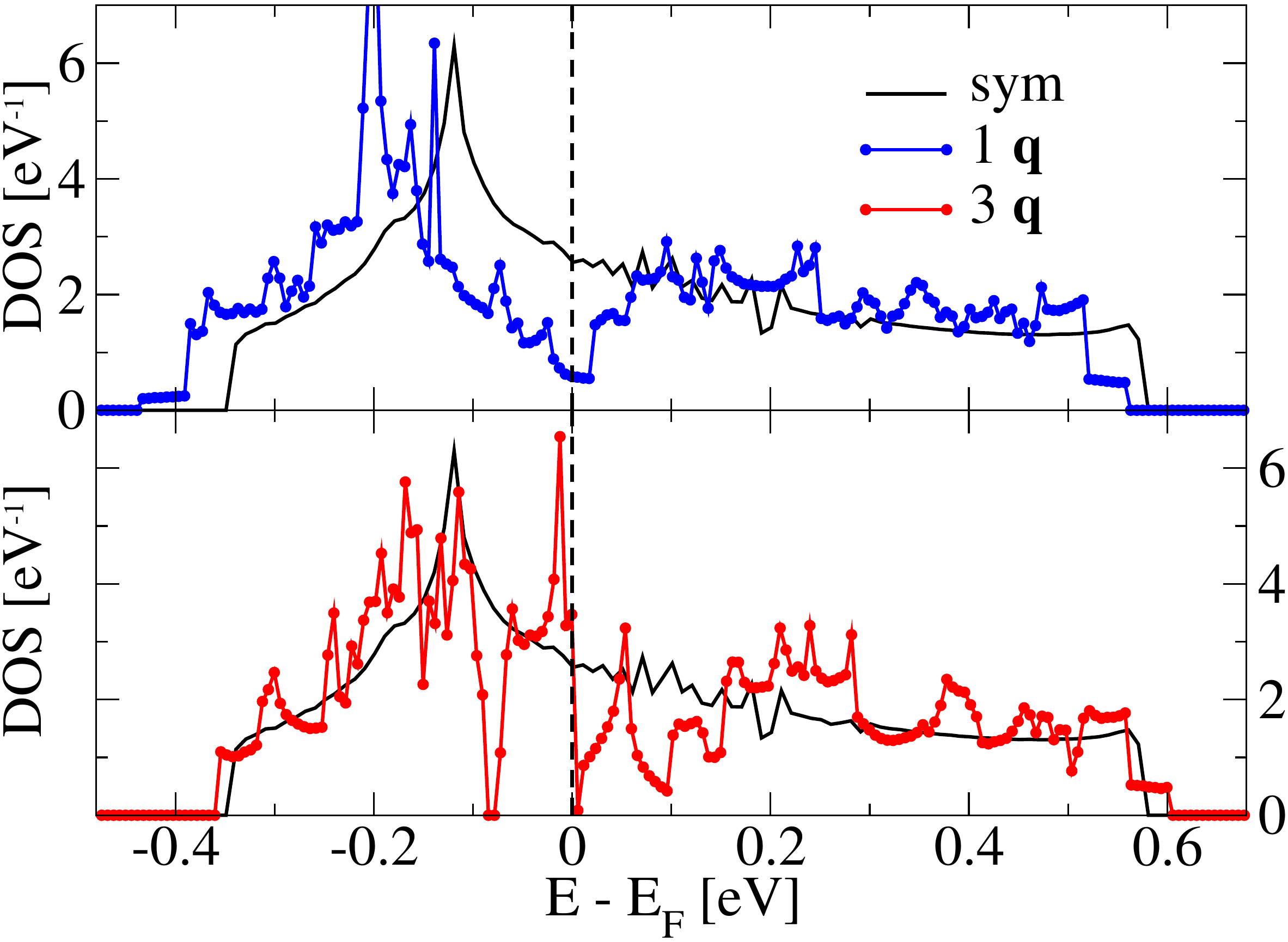}\\
 \caption{Density of the electronic states (DOS) of the non modulated structure
 (in black, denoted by ``sym'') and two representative CDW structures with a
 $4\times4$ periodicity, $1{\mathbf q}$ and $3{\mathbf q}$ (DOS curves for CDW
 structures with a single ordering vector do not differ sensibly). All curves
 are rescaled in formula units for a meaningful comparison between different
 supercells.}
 \label{superdos-fig}
\end{figure}

\begin{figure}[t]
\centering
 \includegraphics[trim = 0 0 0 0,width=0.48\textwidth,clip]{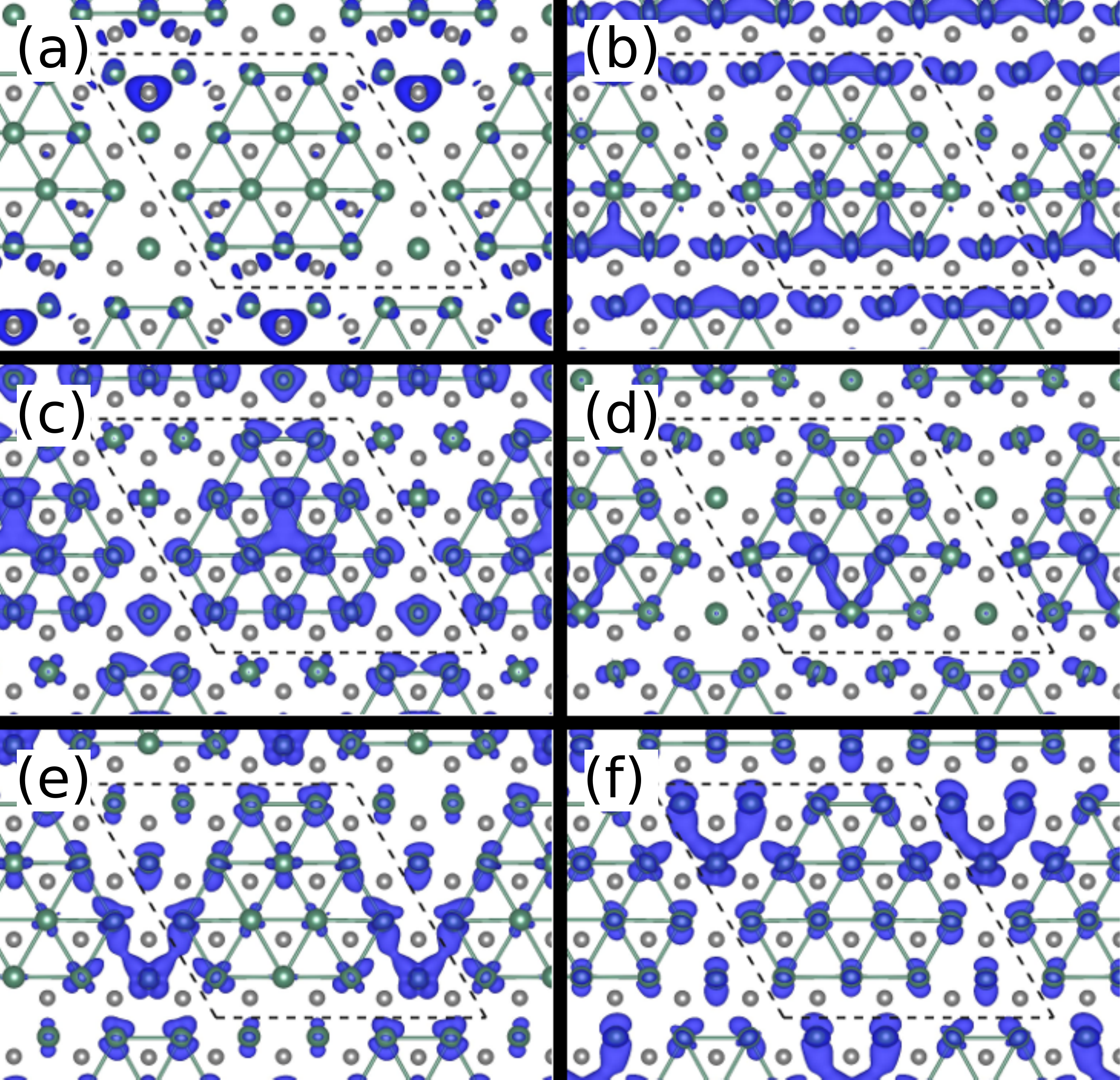}
 \caption{Charge distribution of the $3{\mathbf q}$ CDW integrated over the
 following energy ranges:
 ($-2.65$,$-2.55$) \si{\electronvolt} (a), ($-0.45$,$-0.35$) \si{\electronvolt} (b),
 ($-0.25$,$-0.15$) \si{\electronvolt} (c), ($-0.15$,$-0.05$) \si{\electronvolt} (d),
 ($-0.05$,$+0.05$) \si{\electronvolt} (e), ($+0.05$,$+0.15$) \si{\electronvolt} (f).}
\label{phases3q-fig}
\end{figure}

\begin{figure}[t]
\centering
 \includegraphics[trim = 0 0 0 0,width=0.48\textwidth,clip]{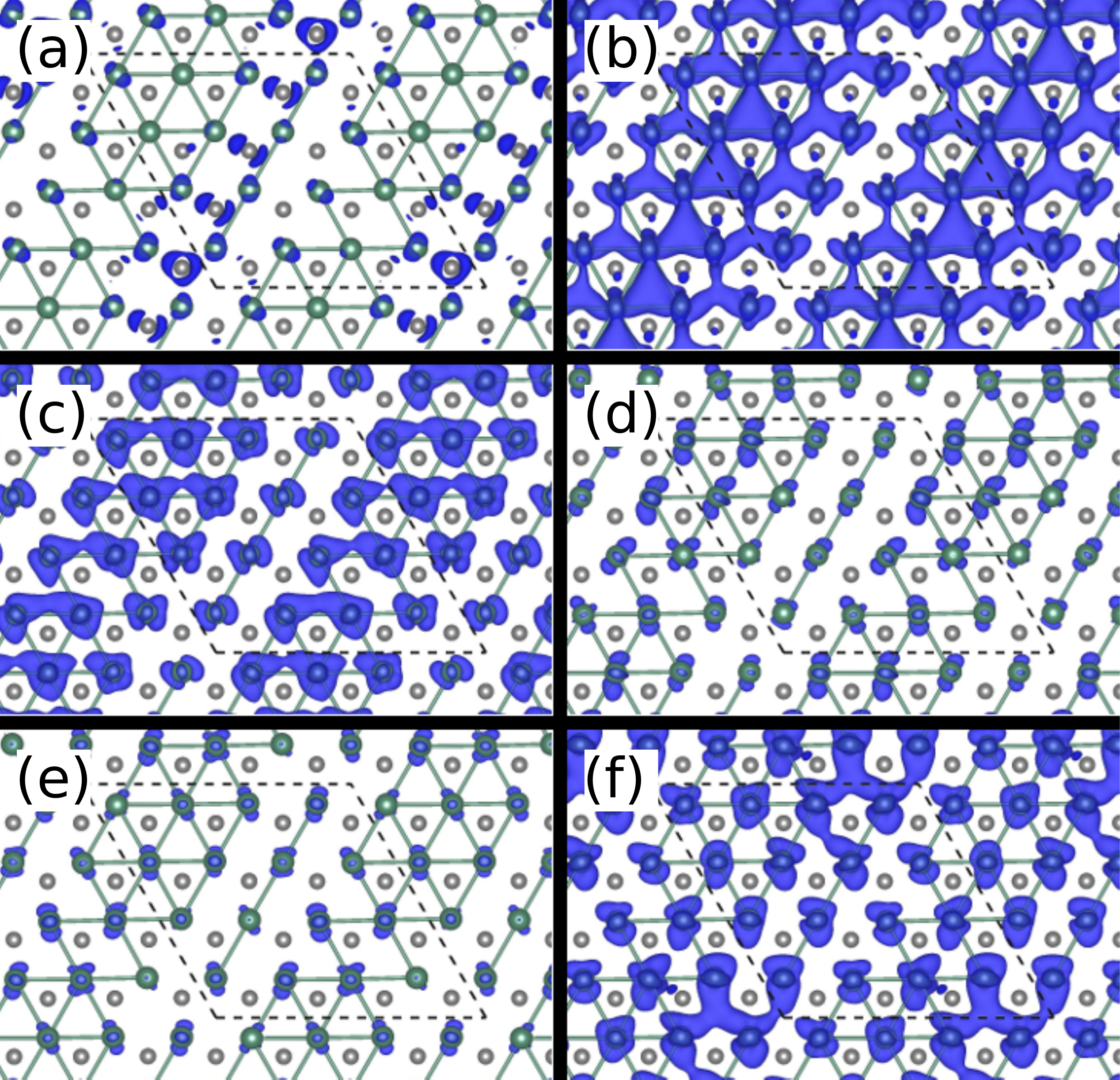}
 \caption{Charge distribution of the $1{\mathbf q}'$ CDW integrated over the
 following energy ranges:
 ($-2.65$,$-2.55$) \si{\electronvolt} (a), ($-0.45$,$-0.35$) \si{\electronvolt} (b),
 ($-0.25$,$-0.15$) \si{\electronvolt} (c), ($-0.15$,$-0.05$) \si{\electronvolt} (d),
 ($-0.05$,$+0.05$) \si{\electronvolt} (e), ($+0.05$,$+0.15$) \si{\electronvolt} (f).}
\label{phases1q-fig}
\end{figure}

  All starting structures for all the supercells considered here have been given a symmetry compatible with a triple ordering vector for the lattice distortion. The relaxed structures
 are found in accordance; however, we find a surprising result in one supercell. A total of three relaxed structures are identified within the $4\times4$ periodicity; two of them
 are characterised by a single ordering vector and referred to as $1\mathbf{q}'$ and $1\mathbf{q}''$ (Figs.\ \ref{4x4strctchg-fig} (a) and (c), respectively); a third one, referred
 to as $3\mathbf{q}$, is characterised by three equivalent ordering vectors, see Fig.~\ref{4x4strctchg-fig} (b); it is energetically unfavoured at the equilibrium lattice constant
 and degenerates into the $1\mathbf{q}'$ ($1\mathbf{q}''$) for compressive (tensile) strain. The wavelength of the $1\mathbf{q}'$ and the $1\mathbf{q}''$ charge distributions is
 \SI{11.82}{\angstrom} ($3.43\times\mathbf{a}_0$, where $\mathbf{a}_0$ is the lattice constant of the unstrained unit cell), \SI{11.96}{\angstrom} ($3.47\times\mathbf{a}_0$) and
 \SI{12.09}{\angstrom} ($3.50\times\mathbf{a}_0$), for compressive strain, no strain and tensile strain respectively, for both phases. These results show a strong connection between
 the phases shown in Fig.~\ref{4x4strctchg-fig} and the stripe phase observed at the surface~\cite{soumyanarayanan-PNAS2013,gao-PNAS2018}. In fact, two features of the modelled
 $4\times4$ CDWs are shown to be in remarkable agreement with experiments: the Fourier transform of the charge distribution -- compare the inset of Fig.~\ref{4x4strctchg-fig} with that
 resulting from STM in figure 1F in ref.\ \onlinecite{gao-PNAS2018} -- and the wavelength of the modulation -- compare our result with the measured wavelength from STM data,
 $\sim$\SI{12}{\angstrom}~\cite{soumyanarayanan-PNAS2013}, $3.5\times\mathbf{a}_0$. Further comparison to these experimental features is provided by our simulated STM plots of
 the $1{\mathbf q}'$ phase, shown in Fig.~\ref{staps4x4-fig}, left column. Additional STM plots, as well as a more extensive discussion, are reported in the Appendix.
 
  The previous analysis of the stripe phase is based on the comparison between our calculations for the single layer (1H) and experimental results for the surface
 (2H). As pointed out in the Introduction, these systems have different characteristics and therefore a certain care is needed when comparing them. To support our
 conclusions, we also performed selected calculations for the bilayer, whose symmetries and physical properties are a better model of the surface. Relevant structures
 with $3\times3$ and $4\times4$ periodicities were simulated, and the latter were found to be more favourable of \SI{0.3}{\eV} with respect to the former (compare
 with Table~\ref{tab-energy}). Further details on these calculations are presented in the Appendix. Finally, it is also important to comment on the
 fact that the stripe phase remains unobserved in single layers (1H), see e.g. ref.\ \onlinecite{ugeda-nphys2016}. This is likely due to unfavourable or insufficient
 epitaxial strain of the substrates where the single layers are deposited, whereas applied~\cite{gao-PNAS2018} and/or accidental strain~\cite{soumyanarayanan-PNAS2013}
 favour such transition. A recent paper by Kim {\em et al.}~\cite{kim_sj-PRB.96.155439} discussed this point in terms of dielectric screening (comparing two different
 substrates), elucidating the role of spin-orbit coupling in the electronic properties and in the CDW formation.
 
  The emergence of a stripe phase under isotropic simulated strain reveals an intrinsic tendency for the CDW towards a lower symmetry, which manifests in the $4\times4$ periodicity and
 remains latent in the $3\times3$ periodicity. In a recent paper~\cite{flicker-PRB.92.201103}, calculations based on a Ginzburg-Landau formalism show that phonon fluctuations suppress
 long-range order in favour of a short-ranged pseudogap phase with a $3{\mathbf q}$ phase, which degenerates into a $1{\mathbf q}$ phase upon uniaxial strain. The authors restrained their
 study to a $3\times3$ CDW. Within $3\times3$ superlattices, with (biaxial) isotropic strain, we cannot find a stripe phase either, and even when a stripe structure is given as an input,
 it degenerates into a triangular CDW (CC or HC, depending on the starting input). All in all, isotropic strain induces only an incipient stripe phase with a $3\times3$ periodicity, see
 Fig.~\ref{3x3striperect-fig}. The HC and CC CDW patches are distinct under tensile strain (c,f) and at the equilibrium lattice constant (b,e), compare the three-spikes ring in the CC
 CDW with the three-fold \textit{fidget-spinner}-like star in the HC CDW, but the HC degenerates into the CC for compressive strain (a,d), where the rings and stars patches are merged.

  On the other hand, allowing the $4\times4$ periodicity, in agreement with the imaginary phonon dispersion, enhances the degrees of freedom and
 widens the ordering parameters space. Such periodicity hosts both single ${\mathbf q}$ and triple ${\mathbf q}$ phases, and therefore is suitable
 to analyse a relation between the electronic reconstruction and the symmetry of the ordering vectors. We consider the electron-phonon coupling as
 the driving mechanism for the CDW, as widely reported in the literature. In a Peierls instability scenario, where the electrons dominate over the
 phonons in driving the transition, a CDW results in a spectral weight loss at the Fermi level, which is also consistent with a weak-coupling. It has
 been already shown~\cite{johannes-PRB.73.205102,silvaguillen-2DMat2016} that it is not the case of NbSe$_2$, where instead the electronic spectral
 weight shifts are more important at a lower energy range, implying that the transition is governed by phonons (note also the presence of electronic
 states below the Nb-derived band and its dependence on the CDW, in ref.\ \onlinecite{silvaguillen-2DMat2016}). Figure \ref{superdos-fig} shows the
 density of the electronic states (DOS) of the single-vector CDWs (denoted both as $1{\mathbf q}$ because their DOS curves have no substantial
 difference), the $3{\mathbf q}$ CDW structure and the undistorted structure. A steep curve, showing increase (depletion) of states below (above)
 the Fermi level, characterises the $3{\mathbf q}$ CDW, whereas the $1{\mathbf q}$ CDW features a depletion of spectral weight all around the Fermi
 level. The $3{\mathbf q}$ CDW displays also a clear and narrow dip at \SI{-0.08}{\eV}. A more detailed analyis of the spectral properties,
 complemented with unfolded band structures, is reported in the Appendix. 
 
  Charge density distributions of the $3{\mathbf q}$ and $1{\mathbf q}$ CDWs in direct space are reported in Figs.\ \ref{phases3q-fig} and \ref{phases1q-fig},
 respectively. These data show a phase change of the electronic modulation across \SI{-0.10}{\eV}, in analogy with ref.\ \onlinecite{arguello-PRB.89.235115}.
 Starting from the Fermi level, see Fig.~\ref{phases3q-fig} (e), the $3{\mathbf q}$ CDW has a $3{\mathbf q}$ symmetry which is maintained down to \SI{-0.25}{\eV},
 see Fig.~\ref{phases3q-fig} (c); below this energy, the symmetry is reduced to a single ${\mathbf q}$. On the other hand, the $1{\mathbf q}$ CDW survives down
 to \SI{-2.6}{\eV}, see Fig.~\ref{phases1q-fig}, pointing to an even stronger phononic involvement in the transition for this phase. The lattice distortions
 induce variations of the DOS at several energy levels, see Fig.~\ref{SI-totdos-fig} in the Appendix, with reference to the $3\times3$ CDW
 ground state, and compare it with ref.\ \onlinecite{silvaguillen-2DMat2016}; around \SI{-2.6}{\eV} and \SI{-1.2}{\eV} large variations occur. Note that the
 symmetry of the charge distribution has particular features depending on the energy range of the electrons. As CDW-induced spectral weight shifts may differ
 in intensity at different energy ranges, STS maps may differ from STM maps; for example, Fig.~\ref{staps4x4-fig} shows that the $3{\mathbf q}$ phase still
 exhibits a three-fold symmetry for all indicated bias voltages in constant-current STM images, where the electronic states are integrated within a corresponding
 energy window (from the Fermi level to the bias voltage).
 
\section{Conclusions}
  In conclusion, we analysed the lattice distortions and their accompanying charge modulations in single layer 1H-NbSe$_2$ under biaxial isotropic strain, finding that a compressive strain
 of $\sim 1 \%$ favours $4\times4$ modulations with a single ${\mathbf q}$ ordering vector. Together with two phases characterised by a single ordering vector, a metastable triple ordering
 vector phase exists. On the other hand, $3\times3$ modulations allow only an incipient $3{\mathbf q}$-$1{\mathbf q}$ transition under strain, but the charge patches of the ground state CDW
 structure assume features from the next most favourable structure. Although the strain in our model may not have been applicable by suitable (e.g. chemically non-reactive) substrates so
 far, the present study offers insights on the intrinsic properties of the single layer NbSe$_2$, which can be useful for future experimental investigations on the manipulation of collective
 excitations in NbSe$_2$ and related transition metal dichalcogenides. Moreover, we observe that the stripe phase observed in NbSe$_2$ thin films is very likely associated to $4\times4$
 modulations, and favoured by compressive strain.\\

\section{Acknowledgements}
  We are grateful to Yunkyu Bang, Poonam Kumari, Dhani Milind Nafday, Erio Tosatti, Stefano de Gironcoli, Ali G.\ Moghaddam, Tristan Cren,
 Hermann Suderow and Han-Woong Yeom for insightful discussions. F.\ C.\ and A.\ A.\ acknowledge financial support from the National Research
 Foundation (NRF) funded by the Ministry of Science of Korea (Grants  No.\ 2016K1A4A01922028, No.\ 2017R1D1A1B03033465, \& No.\ 2019R1H1A2039733); K.\ P.\
 acknowledges support from NRDIO-Hungary project No.\ FK124100. This research was supported by appointments to the JRG program at the APCTP
 through the Science and Technology Promotion Fund and Lottery Fund of the Korean Government, and by the Korean Local Governments --
 Gyeongsangbuk-do province and Pohang City.\newline

\bibliography{biblio.bib}

\newpage

\begin{widetext}
\newpage

\end{widetext}

\setcounter{equation}{0}
\setcounter{figure}{0}
\setcounter{table}{0}
\setcounter{page}{1}
\renewcommand{\theequation}{S\arabic{equation}}
\renewcommand{\thefigure}{S\arabic{figure}}
\renewcommand{\bibnumfmt}[1]{[S#1]}
\renewcommand{\citenumfont}[1]{S#1}

\section*{{\bf  Supplemental material for
\\
Triangular-stripe transition in charge ordered single layer NbSe$_2$}
\\
by}
\subsection*{Fabrizio Cossu, Kriszti\'an Palot\'as, Sagar Sarkar, Igor Di Marco, and Alireza Akbari
}

 \section{Extended DOS and Phase shifts}

\begin{figure}[b]
\centering
 \includegraphics[trim = 0 0 0 0,width=0.48\textwidth,clip]{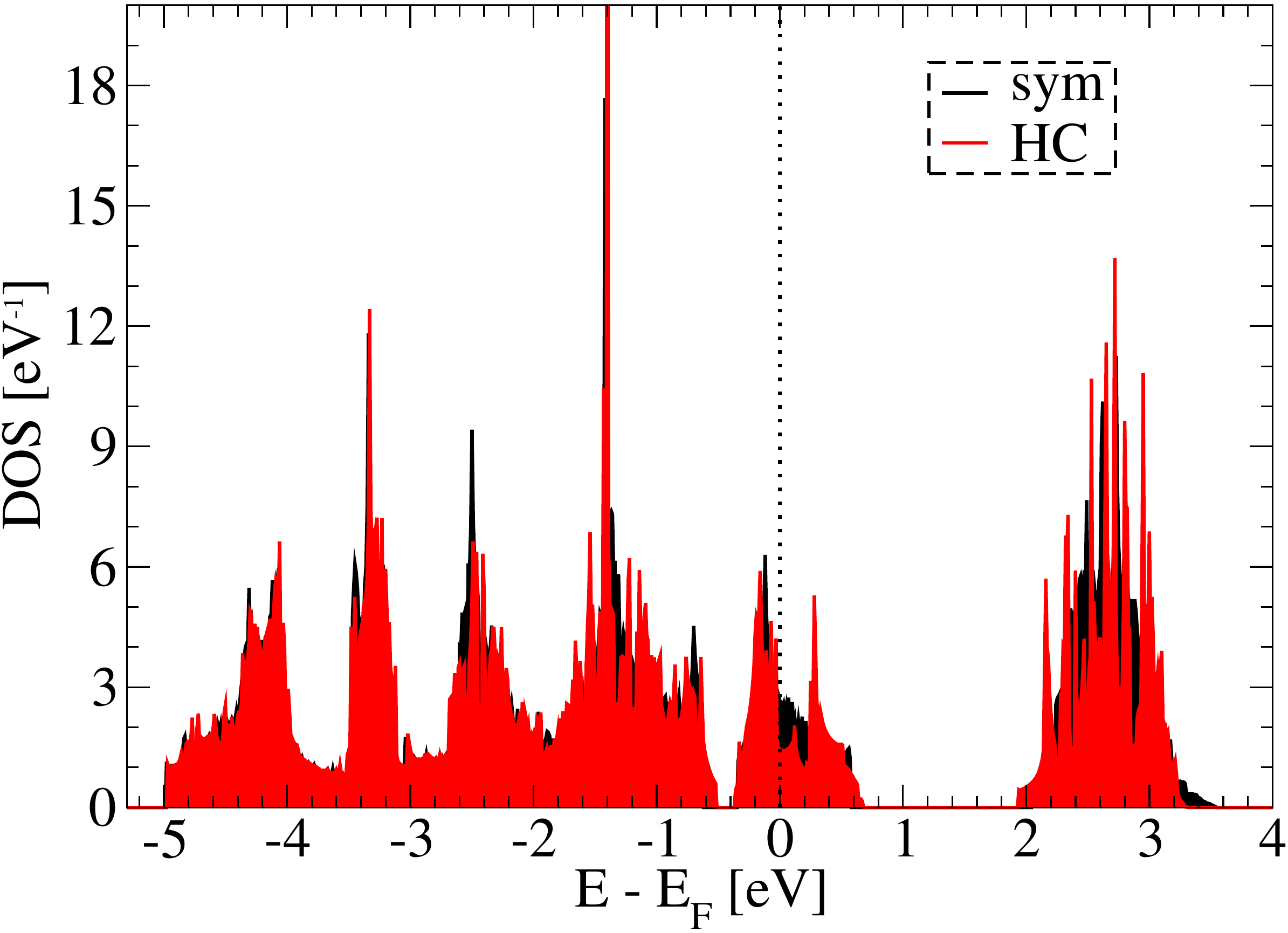}
 \caption{DOS of symmetric and $3\times3$ HC CDW shown in a large range
 to highlight the major changes.}
\label{totdos-fig}
\end{figure}

\begin{figure}[b]
\centering
 \includegraphics[trim = 0 0 0 0,width=0.48\textwidth,clip]{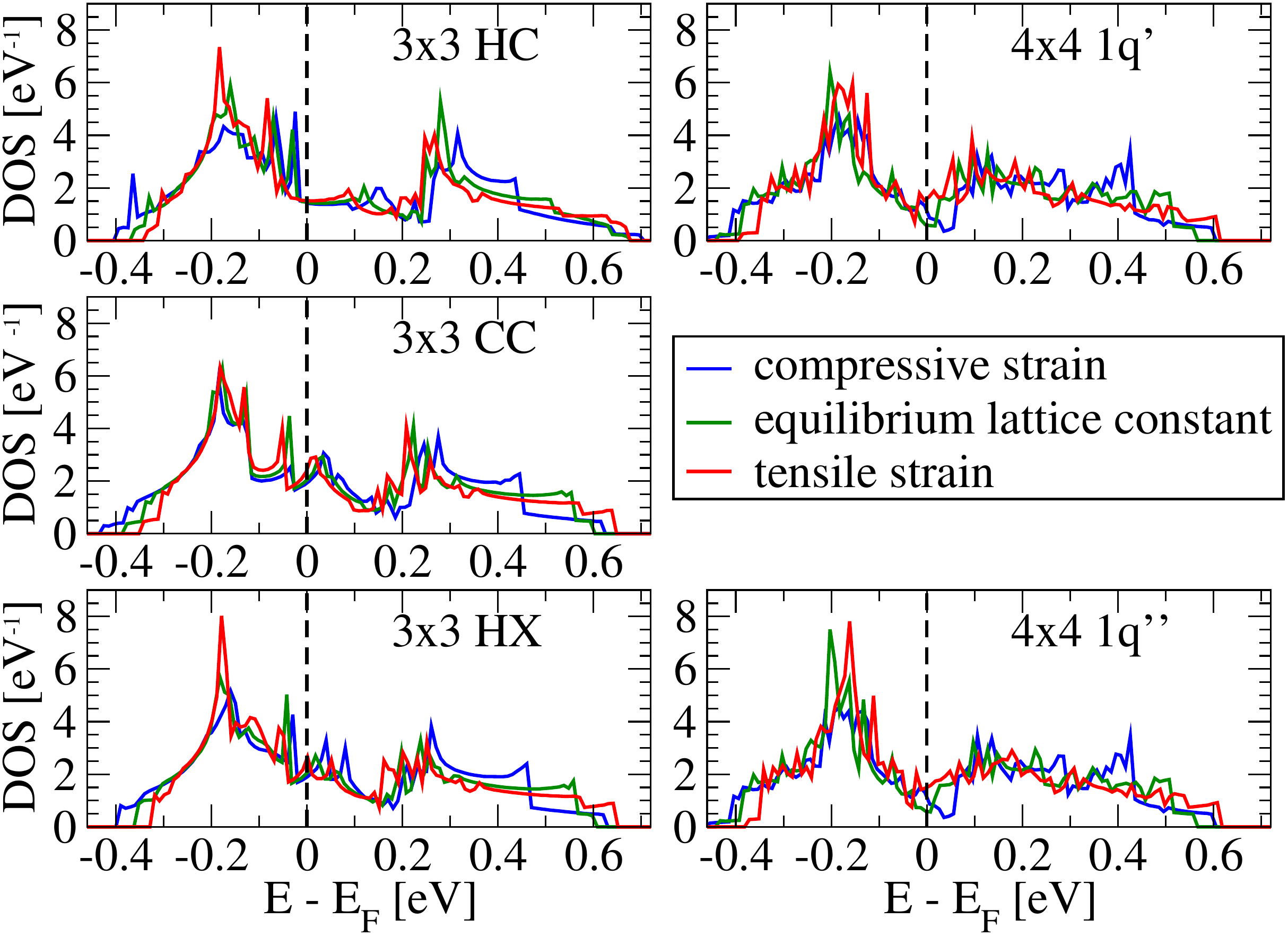}
 \caption{DOS curves for $3\times3$ (left column) and $4\times4$ (right
 column) strained CDWs. The $4\times4$ $3{\mathbf q}$ CDW does not have
 a corresponding structure under strain, and therefore it is not shown.}
\label{straindos-fig}
\end{figure}

\begin{figure}[t]
\centering
 \includegraphics[trim = 0 0 0 0,width=0.48\textwidth,clip]{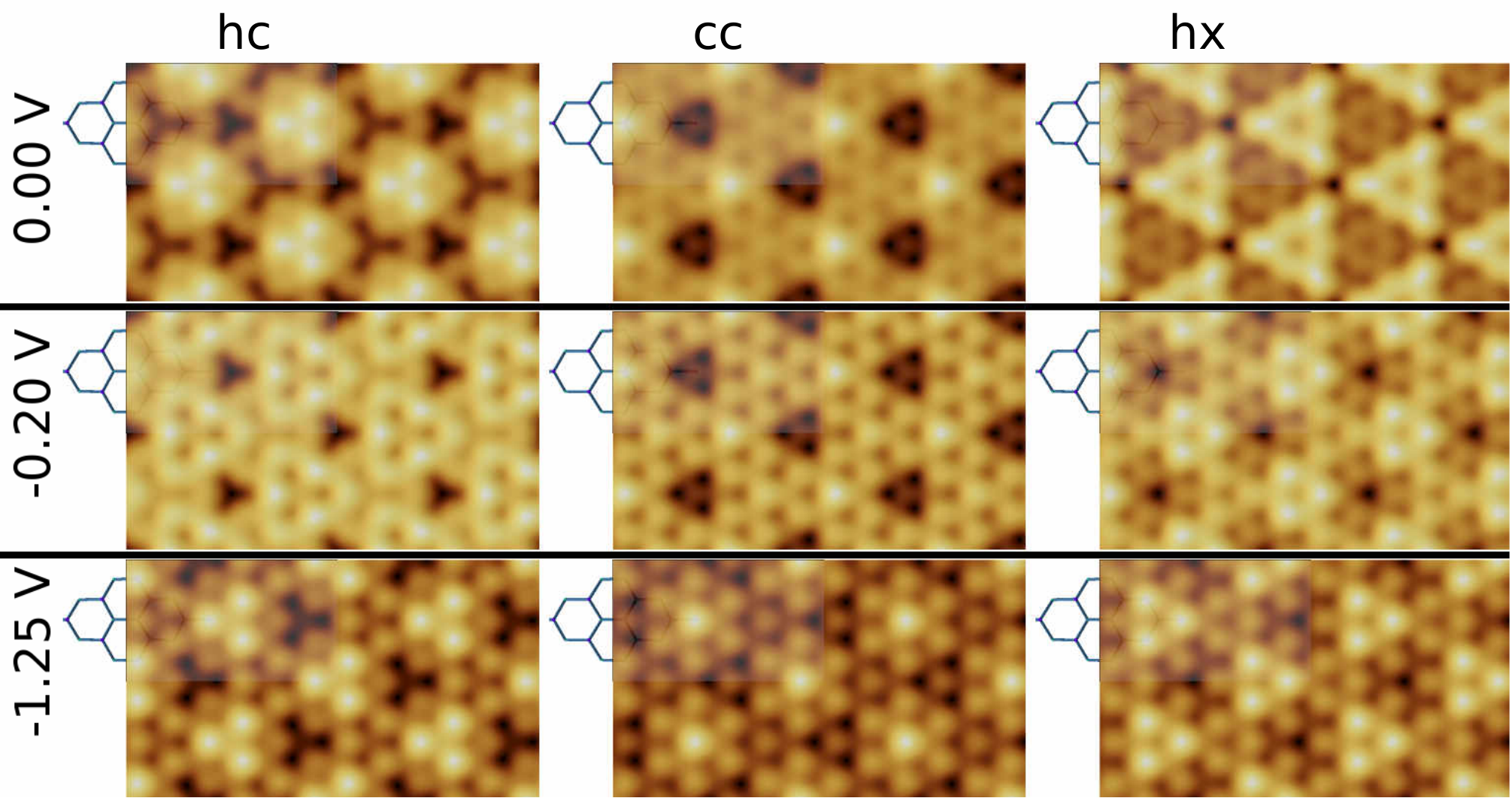}
 \caption{STM maps for the three $3\times3$ HC, CC and HX CDWs,
 respectively from left to right column; raws 1, 2 and 3 stand
 for \SI{0.00}{V}, \SI{-0.20}{V} and \SI{-1.25}{V},
 respectively.}
\label{staps3x3-fig}
\end{figure}

\begin{figure}[b]
\centering
 \includegraphics[trim = 0 0 0 0,width=0.48\textwidth,clip]{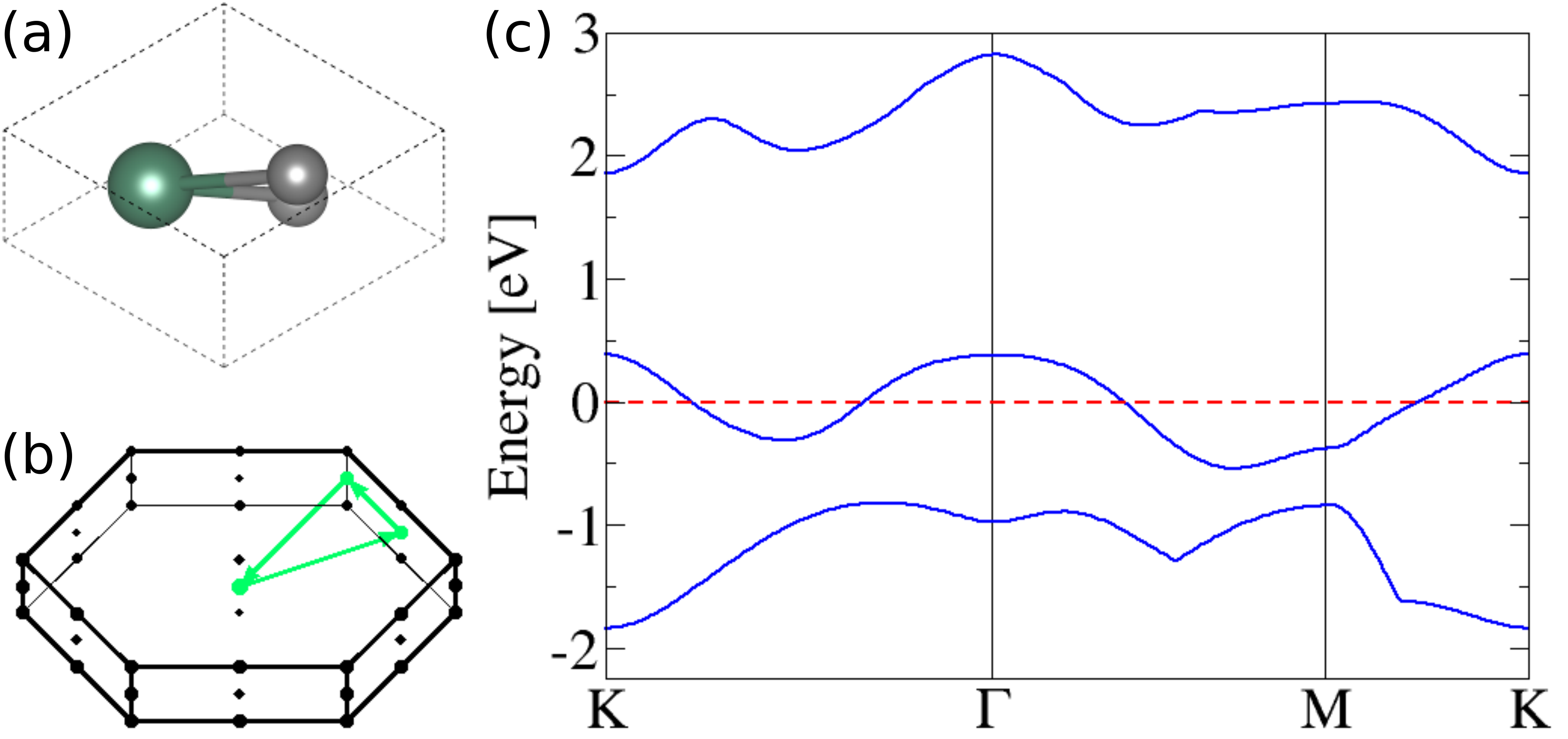}
 \caption{Top-side view of a 1H-NbSe$_2$ unit cell in direct space (a) and
 top-side view of the reciprocal unit cell (b), illustrating the k-path
 selection for the band structure calculation in (c); the band structure
 is computed along the k-path K-$\Gamma$-M-K.}
\label{band-fig}
\end{figure}

\begin{figure*}[bht]
\centering
 \includegraphics[trim = 0 0 0 0,width=0.32\textwidth,clip]{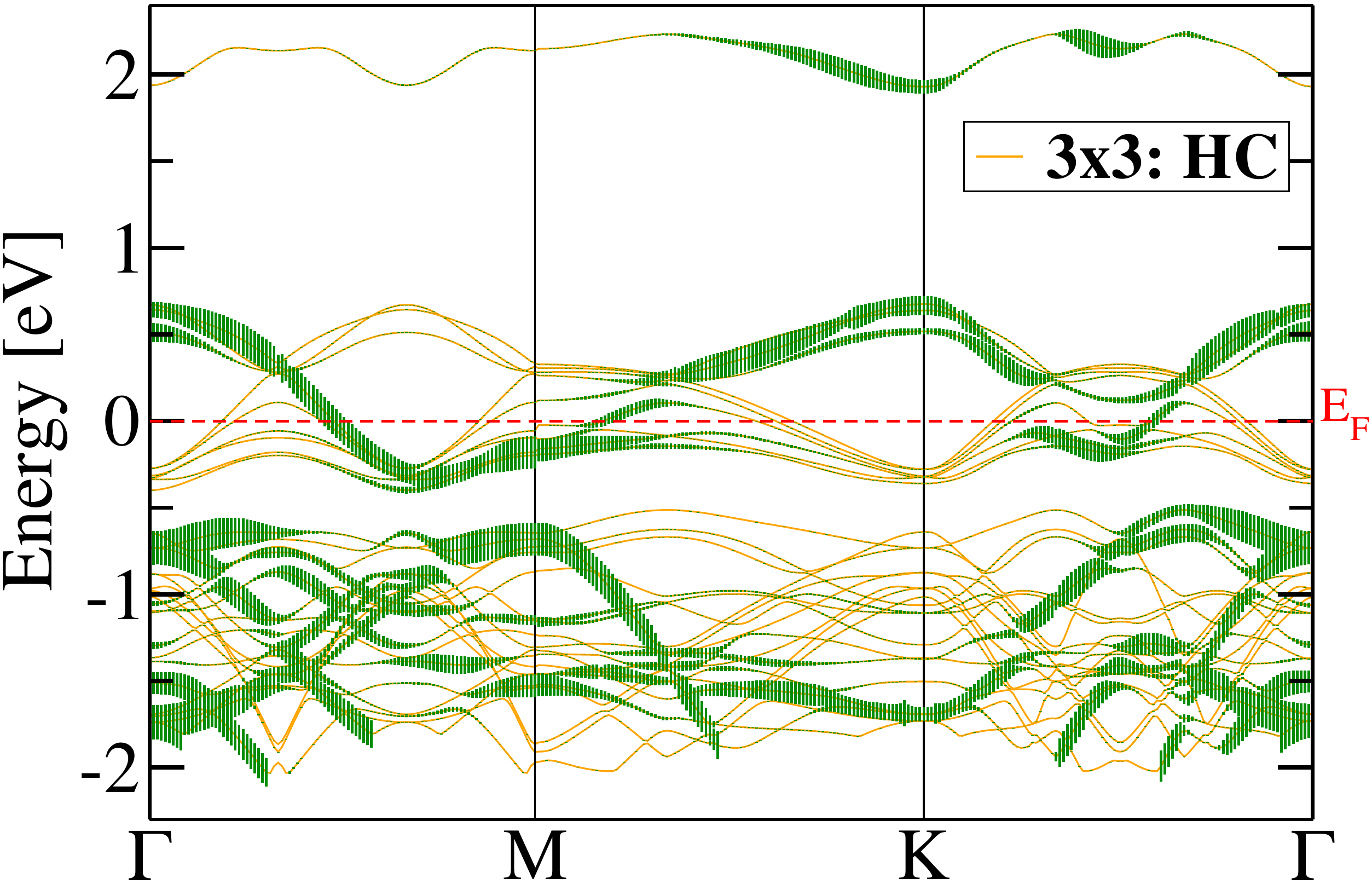}
 \includegraphics[trim = 0 0 0 0,width=0.32\textwidth,clip]{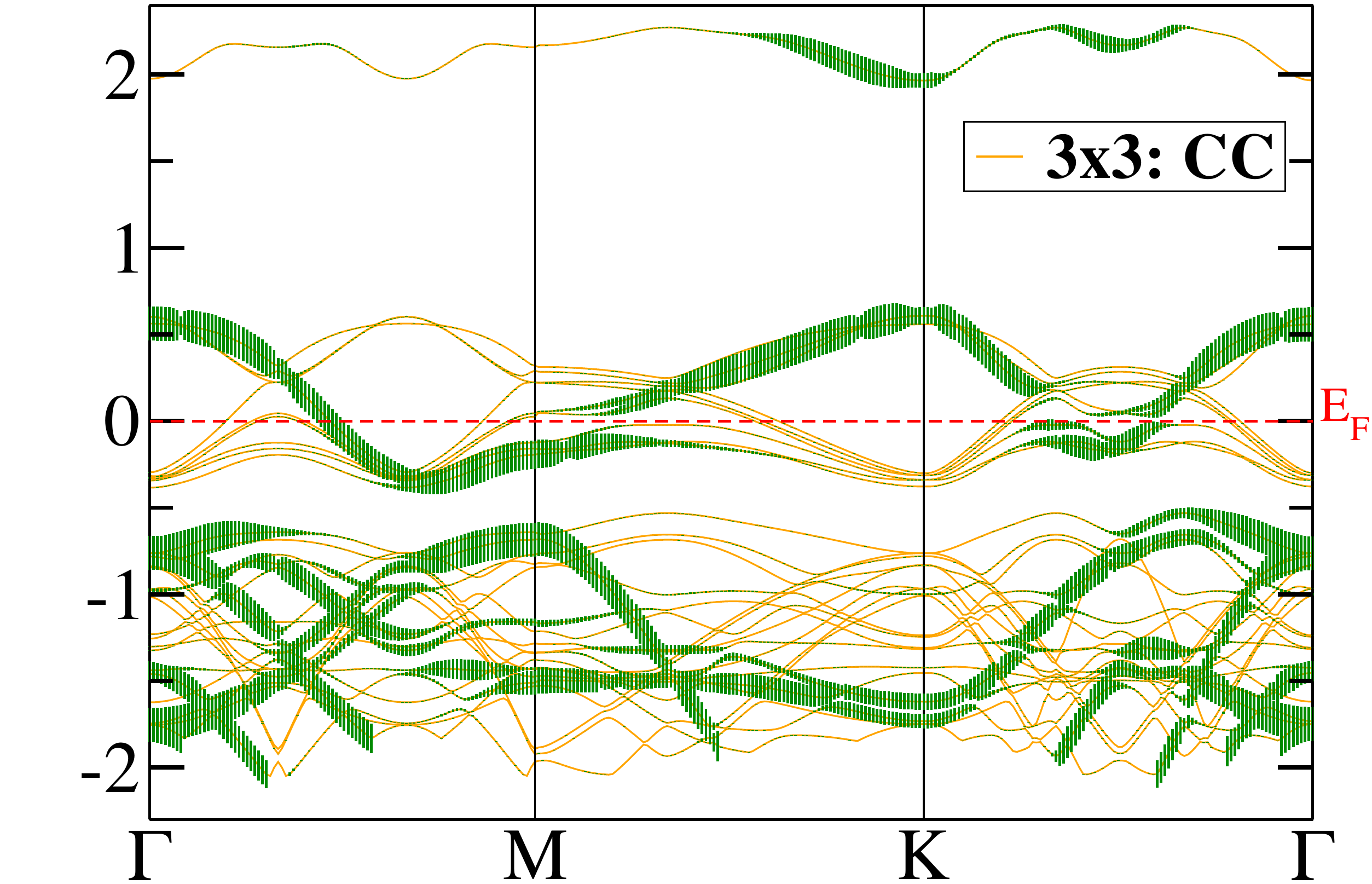}
 \includegraphics[trim = 0 0 0 0,width=0.32\textwidth,clip]{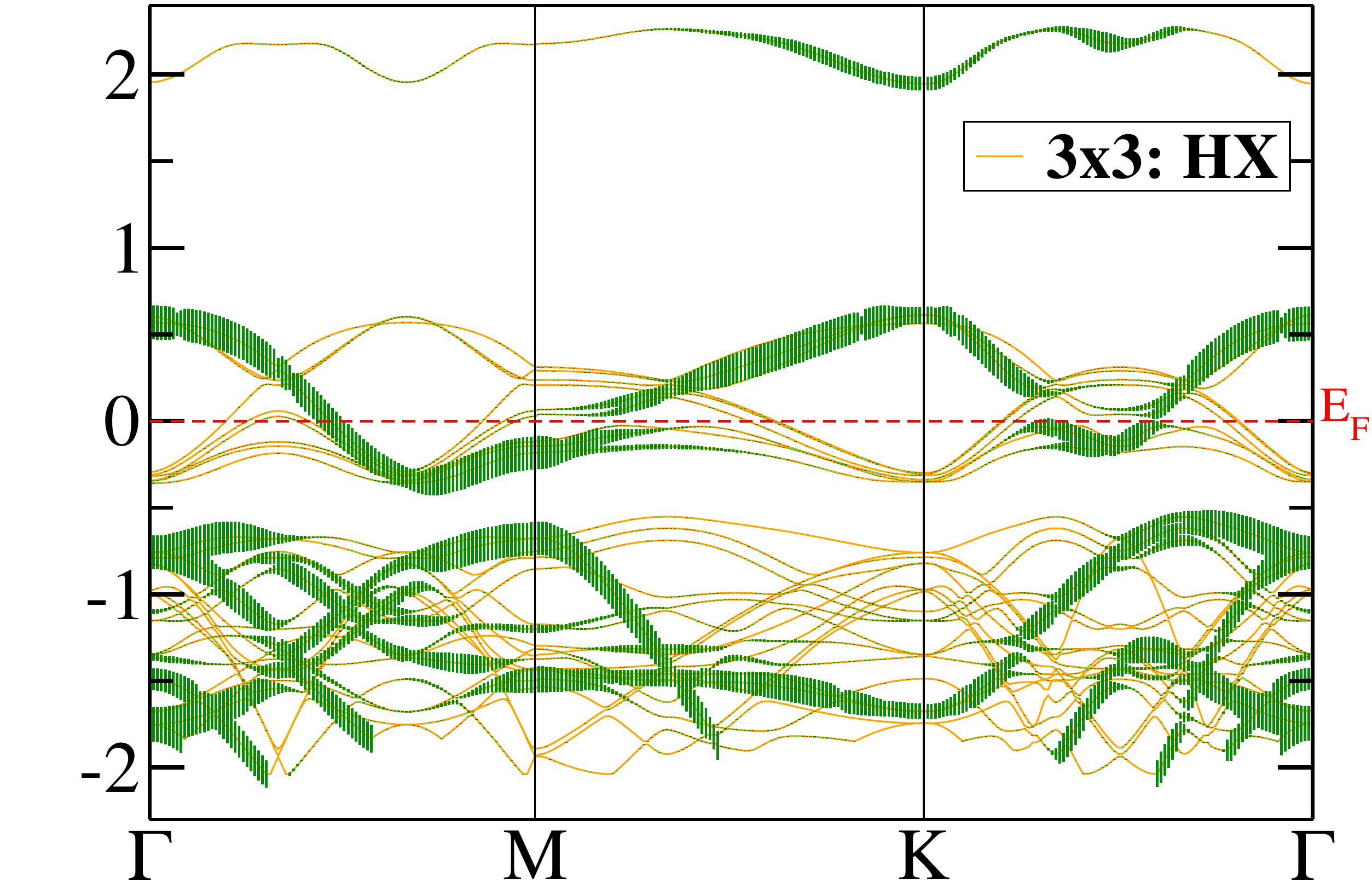}\\
 \includegraphics[trim = 0 0 0 0,width=0.32\textwidth,clip]{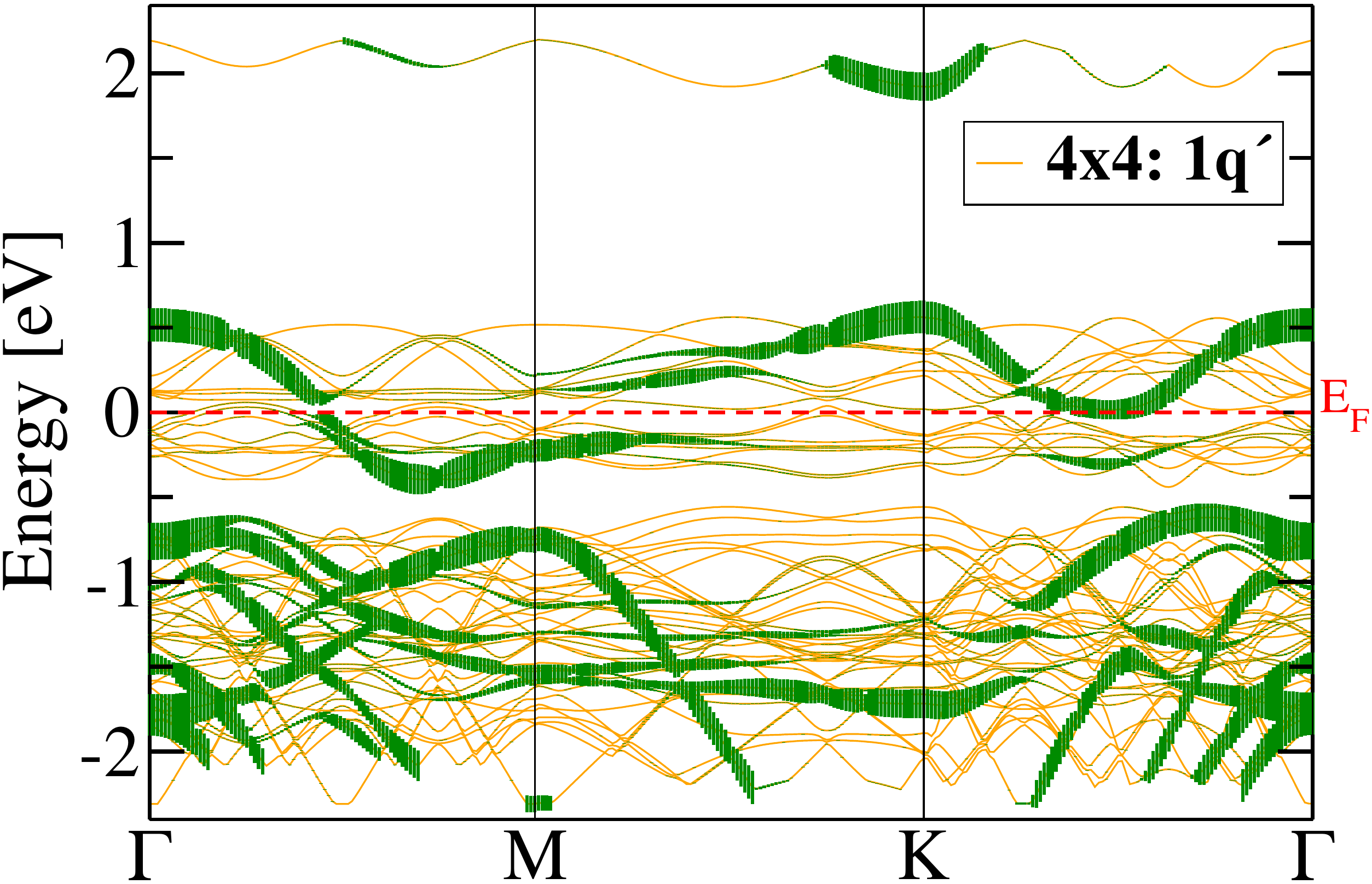}
 \includegraphics[trim = 0 0 0 0,width=0.32\textwidth,clip]{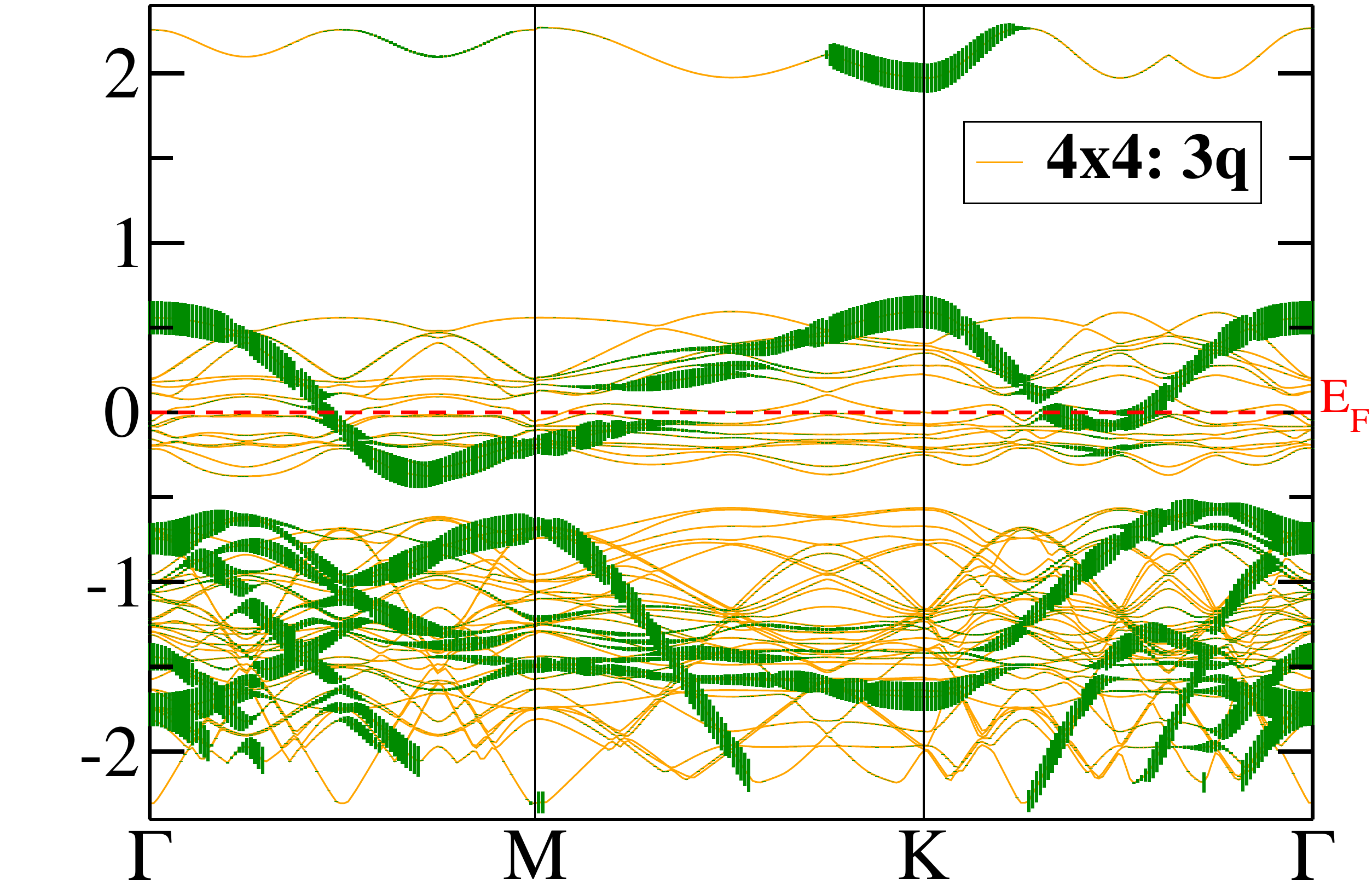}
 \includegraphics[trim = 0 0 0 0,width=0.32\textwidth,clip]{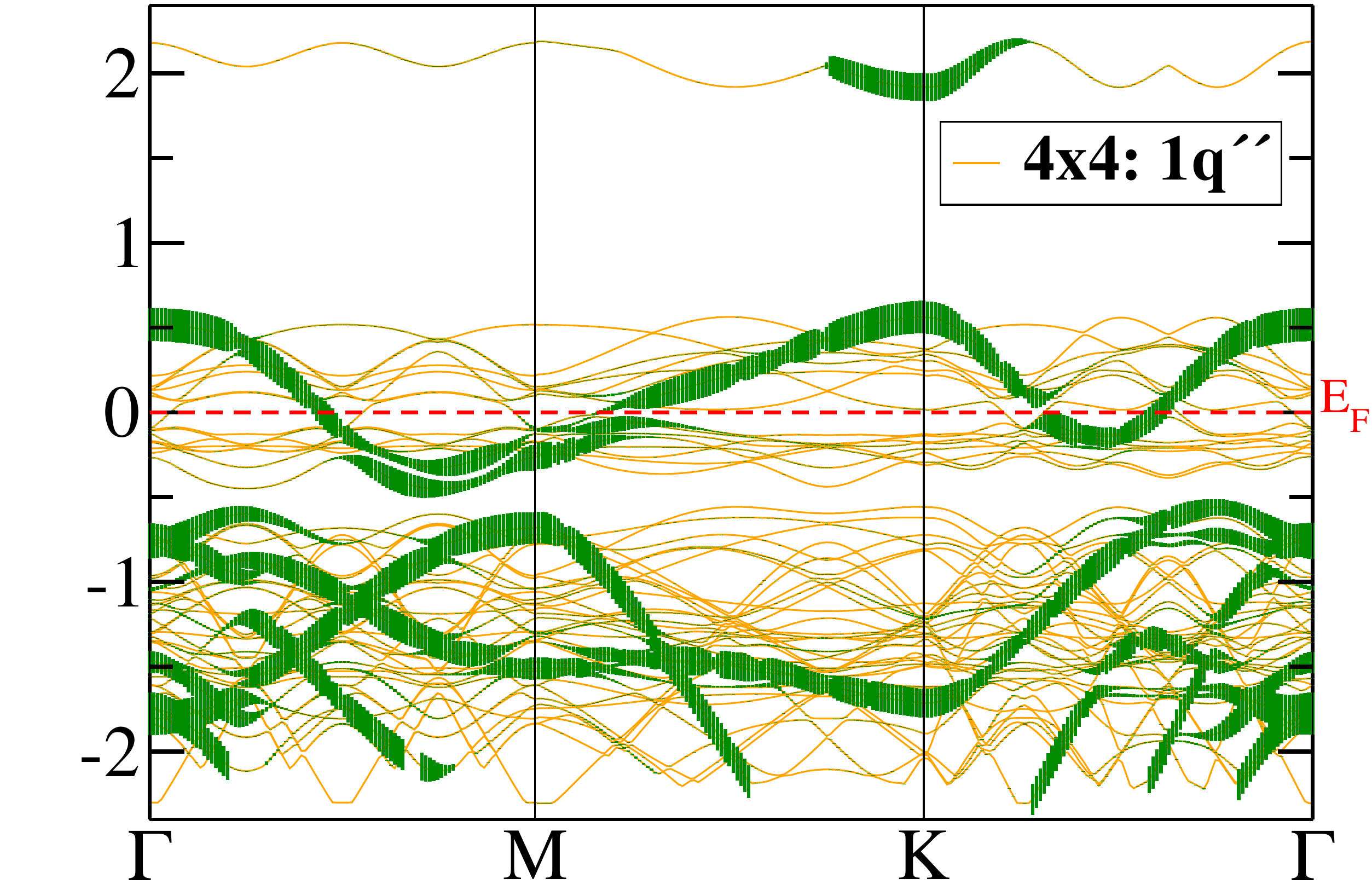}
 \caption{Band structure of the $3\times3$ CDWs (top row) and the $4\times4$
 CDWs (bottom row). Orange lines represent the eigenvalues as computed in the
 folded Brillouin zones (of the relative supercell), whereas the green ticks
 represent the eigenvalues projected onto the unfolded Brillouin zone; their
 thickness are proportional to their spectral weight.}
\label{bunfold-fig}
\end{figure*}

  In the main text, we presented the results for integrated charge density outside the energy range shown in Fig.~\ref{superdos-fig} (main text). For completeness, we report the
 DOS of the $3\times3$ HC CDW and the undistorted structure over an extended energy range in Fig.~\ref{totdos-fig}. This plot provides evidence for the presence of electronic states
 at various ranges and for the regions where major CDW-driven changes occur. We also analyse the phase shifts from Fig.~\ref{phases3q-fig} (main text) in more details. As discussed
 in the main text, the symmetry of this CDW is a $3{\mathbf q}$ around the Fermi level, and down to \SI{-0.25}{\eV}. Within the Nb-derived band, although maintaining a triple ${\mathbf q}$
 ordering vector, a $\pi$ inversion -- Fig.~\ref{phases3q-fig} (c)-(d) (main text) -- and a rigid shift of the charge patches -- Fig.~\ref{phases3q-fig} (e)-(f) (main text)
 -- occur; compare this trend with recent measurements~\cite{SI-arguello-PRB.89.235115}. Conversely, the $1{\mathbf q}$ phases keep the same symmetry, see Fig.~\ref{phases1q-fig} (a)-(f)
 (main text). In the next section, we illustrate the STM maps at different tunnelling voltages. Furthermore, we compare the behaviour of the $3{\mathbf q}$ CDW in the $4\times4$ superlattice
 with the CDWs in the $3\times3$ superlattice. Further consideration is deserved by the strain-induced variations in the DOS, illustrated in Fig.~\ref{straindos-fig}. Important changes at
 the Fermi level are seen especially in the $4\times4$ CDWs, where tensile (compressive) strain tends to increase (decrease) the spectral weight.

 \section{STM/STS}
  Simulations of STM were carried out by employing the BSKAN code~\cite{SI-hofer-PSS2003,SI-palotas-JPCM2005}. The revised Chen method~\cite{SI-mandi-PRB.91.165406} with an
 electronically flat and spatially spherical tip orbital is used, which is equivalent to the Tersoff-Hamann model of electron tunnelling~\cite{SI-tersoff-PRB.31.805}.
 The reported STM images are in constant current mode (with the maxima of the current contours at 5.8 Angstroms for all structures for a better comparability).

  The STM maps of the $3\times3$ and $4\times4$ CDWs are shown in Figs.\ \ref{staps3x3-fig} and \ref{staps4x4-fig} (main text), respectively. For the $3\times3$ CDWs, we find
 a good agreement with existing experimental
 STM data~\cite{SI-soumyanarayanan-PNAS2013,SI-ugeda-nphys2016,SI-arguello-PRB.89.235115,SI-chatterjee-ncomm2015,SI-gye_gc-PRL.122.016403,SI-gao-PNAS2018,SI-nakata-NPJ2DMA2018}. First, we point out that
 it is well known that the tip-sample distance can have a major influence on the observed STM contrasts, and a complete contrast reversal is also possible, see e.g. ref.\
 \onlinecite{SI-palotas-PRB.86.235415}. This effect is crucial for CDWs in hetero-layered structures such as transition metal dichalcogenides as well, where the more charged regions can
 be imaged as protrusions or depressions depending on the tip and tunneling conditions. In our case, modulations can result from a superposition of contribution from the transition
 metal layer and the chalcogen layer. As illustrated in figures 9 and 1 of ref.\ \cite{SI-cossu-PRB.98.195419}, the Se-Se bond patterns of CC and HX have the same symmetry as the Nb-Nb
 bond patterns of HX and CC, respectively. Therefore, analysing the STM maps and the charge distribution, some care has to be taken. As our set of examples show, the observed STM
 contrasts may depend dramatically on the bias voltage and on the tip-sample distance (not shown), e.g., all observed STM contrasts are inverted at -1.25 V with respect to the other
 considered voltages in both Figs.\ \ref{staps3x3-fig} and \ref{staps4x4-fig} (main text). In some cases, the contrast, see e.g. Fig.~\ref{staps3x3-fig}, is completely inverted
 with respect to what it is expected from the visualised isosurface charge, shown e.g. in Fig.~\ref{3x3striperect-fig} (main text) or in previous \textit{ab-initio}
 studies~\cite{SI-cossu-PRB.98.195419}. The reasons for this puzzling trend are the different electronic states contributing to the tunneling current at different bias voltages, and the
 variation of the decay of the electron wave functions of different symmetries in the vacuum. Nevertheless, a good agreement with a measured phase shift~\cite{SI-arguello-PRB.89.235115}
 is found. The STM maps for two of the $4\times4$ CDWs, $1{\mathbf q}'$ and $3{\mathbf q}$ are shown in Fig.~\ref{staps4x4-fig} (main text). As opposed to the integrated charge
 density around specific energy values, where the symmetry of the $3{\mathbf q}$ is not maintained below a certain energy, these STM maps show that the three-fold symmetry is maintained
 for integrated states from the Fermi level below \SI{-0.4}{\V}, pointing to a stronger influence of the Nb-derived band in determining the symmetry of the charge modulation. A phase
 shift of the $1{\mathbf q}'$ phase, more clear than in the charge plots shown in Fig.~\ref{phases1q-fig} (main text) is observed; the same phase shows a clearer stripe character
 as a consistent voltage is considered. Regarding the comparison between periodicities, the STM maps for the $3{\mathbf q}$ resemble some characteristic feature of the $3\times3$ CDWs.

 \section{Band structures}
  In view of further exploration of the stripe phase and the $4\times4$ vs $3\times3$ periodicity, we analyse the band structures for the $3\times3$ and $4\times4$ CDWs, with
 reference to Fig.~\ref{bunfold-fig}, comparing it with the band structure of a unit cell, shown in Fig.~\ref{band-fig}. The calculated band structure of the supercells were
 unfolded back to the primitive cell Brillouin zone to get an effective band structure of the systems~\cite{SI-popescu-PRL.104.236403}. This is done following the method of Popescu
 and Zunger~\cite{SI-popescu-PRB.85.085201}, which is based on calculating the spectral weights of the supercell eigenvalues and constructing the spectral function. The orange lines
 in Figs.\ \ref{bunfold-fig} correspond to the eigenvalues within the Brillouin zone of the supercell (folded), whereas the green curves represent the eigenvalues projected onto
 the unfolded Brillouin zone with a broadening proportional to their spectral weight. The Nb-derived band crossing the Fermi level features band splittings around M and along
 K$\Gamma$, but the latter is much less pronounced for the $4\times4$ CDWs. The valence band below features a splitting along K$\Gamma$, MK and $\Gamma$M, accompanied to a
 plethora of valence bands arising from the lattice distortions. Compared to angle resolved photoemission spectroscopic
 measurements~\cite{SI-bawden-ncomm2016,SI-nakata-NPGAM2016,SI-ugeda-nphys2016,SI-nakata-NPJ2DMA2018}, these bands look more detailed. The conduction band splitting along $\Gamma$M and MK
 are in agreement; the lower valence band features a two-fold bifurcation at $\sim$ \SI{-1}{\eV} along $\Gamma$M and especially $\Gamma$K; recent experiments~\cite{SI-nakata-NPJ2DMA2018}
 detect such feature along $\Gamma$K only, as a result of spin-orbit interaction -- which in our calculations is not taken into account.

\begin{figure*}[t]
\centering
 \includegraphics[trim = 0 0 0 0,width=0.78\textwidth,clip]{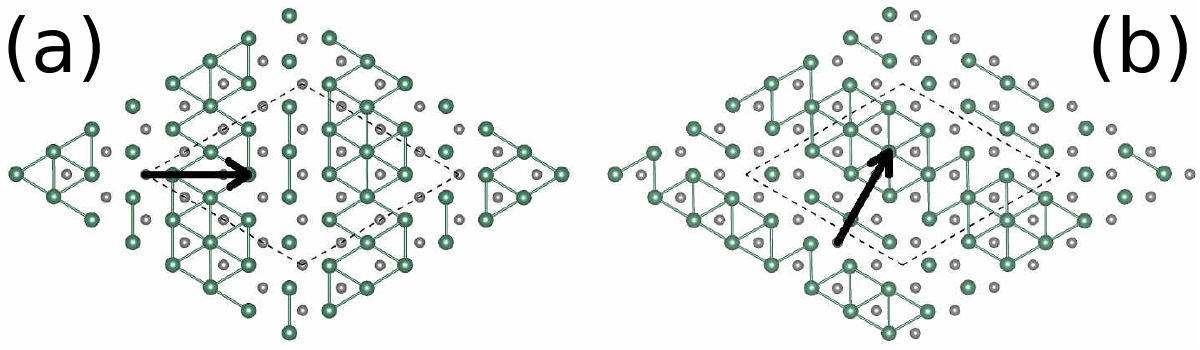}
 \caption{Structures of upper (a) and lower (b) layers of a $4\times4$
 bilayer shown separately and juxtaposed. In (a) and (b), black arrows
 depict the ${\mathbf q}$ ordering vectors.}
\label{bilayers-fig}
\end{figure*}

 \section{A hint to the properties of a bilayer}
  Throughout this manuscript, we presented a model for a stripe phase in single layer 1H-NbSe$_2$. Although the stripe phase was not observed for single layers, it was observed for
 thin films at the surface. As mentioned in the main text, the strain a substrate can apply on a single layer may not be sufficient, or it may be opposite in sign, for the stripe
 phase to occur. In order to strengthen our argument, we computed the phononic spectra of a bilayer 2H-NbSe$_2$, using a k-mesh of $40\times40\times2$ for the wavefunctions and a
 q-mesh of $8\times8\times1$ for the force constants. Due to the centrosymmetry of the 2H-bilayer, instabilities occur along the $\Gamma$M line (around 2/3). We modelled bilayers
 with $3\times3$ and $4\times4$ periodicities, taking few examples from the single layer case, finding that the energy difference between $4\times4$ and $3\times3$ lattice distortions
 (\SI{0.3}{meV} per unit cell in favour of the $4\times4$ periodicity) is in line with the above reported results for the single layer. Although we have not tested all possible
 configurations, these results suggest that the ordering between $3\times3$ and $4\times4$ is not highly altered by the symmetry; however, defining a clear trend is not within
 the scope of the present work. The relaxed structures for the $3\times3$ periodicity maintain the same geometry they have in the single layer, whereas the $3{\mathbf q}$ interfaced
 with a $1{\mathbf q}$ relaxes to a $1{\mathbf q}$ structure; note that the ordering vectors of the two interfaced layers are twisted by $\pi/3$ with respect to each other, see
 Fig.~\ref{bilayers-fig}. The $3\times3$ bilayers are not shown because there is no new information about the structure with respect to the single layers.



\end{document}